\documentclass[a4paper, 12pt]{article}
\usepackage[margin=25mm]{geometry}
\usepackage{floatflt}
\usepackage[utf8]{inputenc}
\usepackage{csvsimple}
\usepackage{fp}
\usepackage{colortbl}
\usepackage{amsmath}
\usepackage{amsfonts}
\usepackage{amssymb}
\usepackage{graphicx}
\usepackage{xcolor}
\usepackage[nottoc, notlof, notlot]{tocbibind}
\usepackage{makeidx}
\usepackage{hyperref}
\usepackage{diagbox}
\usepackage{array,multirow,makecell}
\usepackage{subcaption}
\usepackage{setspace} 
\newcommand{\integ}[2]{\displaystyle{\int_{#1}^{#2}}}
\newcommand{\vitfull}[0]{{\bf u}_{full}}
\newcommand{\vitPOD}[0]{{\bf u}_{POD}^N}

\newcommand{\dx}[0]{{\bf dx}}
\newcommand{\fract}[2]{\displaystyle{\frac{#1}{#2}}}

\newcommand{\porosity}[0]{\varepsilon}
\newcommand{\base}[1]{{\varphi}_{#1}}
\newcommand{\basevit}[1]{\boldsymbol{\varphi}_{#1}^u}
\newcommand{\esp}{L^2(\Omega)}
\newcommand{\ensemble}[0]{{\cal{E}}}
\newcommand{\nbcliche}[0]{N_s}
\newcommand{\taillebase}[0]{m}
\newcommand{\moyth}[1]{<#1>}
\newcommand{\scalth}[2]{\big(#1,#2\big)}
\newcommand{\param}[0]{\boldsymbol{\gamma}}
\newcommand{\paramB}[0]{\boldsymbol{\gamma}}
\newcommand{\fonc}[0]{{w}}
\newcommand{\varspace}[0]{{\bf{x}}}

\newcommand{\ensemblebase}[0]{\Phi}
\newcommand{\dis}[1]{\displaystyle{#1}}
\providecommand{\keywords}[1]
{
	\small	
	\textbf{\textit{Keywords---}} #1
}
\title{Solving flows in porous media with a POD-Galerkin reduced order model coupled with multilayer perceptron}
\author{C. Allery, C. Béghein, C. Dubot, F. Dubot}
\begin{document}
\maketitle
\begin{center}
	LaSIE, UMR CNRS 7356, La Rochelle Université\\
	Av. Michel Crépeau, 17042 La Rochelle Cedex 1, France\\
	\{callery, cbeghein\}@univ-lr.fr\\
	dubotclaire@yahoo.fr\\
	fdubot@outlook.com\\
\end{center}
\begin{abstract}
This paper deals with the numerical modeling of flow around and through a porous obstacle by a reduced order model (ROM) obtained by Galerkin projection of the Navier-Stokes equations onto a Proper Orthogonal Decomposition (POD) reduced basis. In the few existing works dealing with model reduction techniques applied to flows in porous media, flows were described by Darcy's law and the non linear Forchheimer term was neglected. This last term cannot be expressed in reduced form during the Galerkin projection phase. Indeeed, at each new time step, the norm of the velocity needs to be recalculated and projected, which significantly increases the computational cost, rendering the reduced model inefficient. To overcome this difficulty, we propose to model the projected Forchheimer term with artificial neural networks. Moreover in order to build a stable ROM, the influence of unresolved modes and pressure variations are also modeled using a neural network. Instead of separately modeling each term, these terms were combined into a single residual, which was modeled using the multilayer perceptron method (MLP). The validation of this approach was carried out for laminar flow past a porous obstacle in an unconfined channel. The proposed ROM coupled with MLP approach is able to accurately predict the dynamics of the flow while the standard ROM yields wrong results. Moreover, the ROM MLP method improves the prediction of flow for Reynols number that are not included in the sampling and for times longer than sampling times.
\end{abstract}
\section{Introduction}
Flows in porous media are encountered in various applications (water infiltration in the soil, flows in biomaterials, geothermal energy production, etc.). More particularly, flows past porous obstacles have attracted attention in the litterature for a long time \cite{Valipour}\cite{Yu1}\cite{Yu2} due to their practical applications (flows around buildings, bridges, etc.).
Numerical simulation of flows in porous media using classical approaches, such as the finite element method or the finite volume method, requires long computation times and significant storage capabilities. These methods cannot be effectively used in processes where the problem needs to be solved multiple times, such as real-time control or geometric optimization during a design phase. To overcome these difficulties, model reduction techniques can be employed, which involve constructing a reduced spatial basis able to reproduce the flow using a minimal number of basis vectors. Once this basis is constructed, the temporal and/or parametric dynamics are obtained through a reduced model.

Numerous approaches exist for constructing this reduced basis: a posteriori linear approaches such as Proper Orthogonal Decomposition (POD) method \cite{Berkooz1993}\cite{Sirovich87}, balanced truncation method \cite{Rowley2005}, Reduced Basis method \cite{Buffa2012}\cite{Quarteroni2015}, or Dynamic Mode Decomposition (DMD) method \cite{Schmid2010}; a priori linear approaches such as A Priori Reduction (APR) method \cite{Ryckelynck2005}\cite{Verdon2011} or Proper Generalized Decomposition (PGD) method \cite{Ammar2007}\cite{Dumon2011}; and nonlinear approaches such as kernel PCA \cite{Scholkopf98} method, the Locally Linear Embedding (LLE) method \cite{Roweis2000}, or the ISOMAP method \cite{Tenenbaum2000}.

The most widely used method is the POD method, which involves generating an optimal reduced spatial basis from a set of snapshots, where the first vectors capture the majority of the information contained in the initial datas. Once this basis is obtained, the governing equations of the problem are projected, using Galerkin projection, onto each element of this truncated basis to obtain a system of ordinary differential equations of small size that can be solved quickly. The POD-ROM Galerkin approach has been extensively applied in fluid mechanics \cite[\dots]{Allery2005, Beghein2014, Semaan2016, Stabile2017, Stabile2018, Girfoglio2022, Buoso2022}, particularly for real-time control \cite[\dots]{Ravindran1997, Tallet2016, Nagarajan2018, Oulghelou2021_JCP}. However, very few studies have been conducted on porous media. In this scope,  the approach has been used to simulate :  the non-linear miscible viscous fingering in a 2D porous medium \cite{Chaturantabut2011}, the solution of transport of passive scalars in homogeneous and heterogeneous porous media  \cite{RizzO2018}, two-phase flows in porous media \cite{Li2019}, the steady-state flow in fractured porous media\cite{Li2020}, or the single‑phase compressible flow in porous media \cite{Li2021}. Nevertheless, in these works, flows  were described by Darcy’s law, and the nonlinear inertial term of Forcheimer was neglected. Indeed, this term ${\bf F_{for}}=\alpha \|{\bf u }\|{\bf u}$, where ${\bf u}$ is the fluid velocity, cannot be expressed in reduced form during the Galerkin projection phase. At each new time step, the norm $\|{\bf u }\|$ needs to be recalculated and projected, which significantly increases the computational cost, rendering the reduced model inefficient. In \cite{German2021}, the projection of this term onto the reduced basis, is treated using the Discrete Empirical Interpolation Method (DEIM).  In this paper, to overcome this difficulty, we propose modeling these projected terms using artificial neural network (ANN), where the inputs are the coefficients of the reduced model. 

Neural network approaches coupled with reduced-order models obtained through POD are increasingly being used, in particular, to construct data-driven models. For instance, in \cite{Hesthaven2018}, Hesthaven and Ubbiali propose a non-intrusive reduced basis method for parametrized steady-state partial differential equations. The method extracts a reduced basis from a collection of high-fidelity solutions via the POD and employs ANN, particularly multi-layer perceptrons (MLP), to approximate the coefficients of the reduced model. Pawar et al. \cite{Pawar2019} have built a surrogate model for complex fluid flow utilizing POD to
compress data and deep neural network to forecast dynamics of the system. Similar approach has been developped to predict unsteady airfoil flow variables of interest for aerodynamic applications\cite{DiasRibeiro2023}. In \cite{Demo2021} authors build a probabilistic response surface of these coefficients using the Gaussian process regression. The same framework is used to solve linear poroelasticity problems in heterogeneous porous media \cite{Kadeethum2021}. During the online phase, the trained artificial neural networks are evaluated to obtain coefficients corresponding to new values of uncertain parameters (material properties, boundary conditions, or geometric characteristic). In \cite{Ahmed2021b}, a
plain multilayer autoencoder \cite{Sondak2021} is constructed  to generate a nonlinear mapping between these POD coefficients and a latent space constructed with only a few parameters in the bottleneck layer. This approach reduces the computational cost of the training process since it involves significantly fewer trainable parameters. The potential of this approach, namely Nonlinear POD, is shown on Marsigli convection dominated flow problem. In \cite{Akbari2022} the approach is extended to Rayleigh Bénard convection problem. One of the remarks often made for these data driven methods concerns their failure to take physics into account.
Thereby, recently, Physics Informed Neural Network (PINN) has been developed to solve partial differential equations  \cite{Raissi2019}. The idea is to incorporate the residual of the PDE as part of the loss function, to make the network solution underlying physics. Chen et al. \cite{Chen2021} proposed a reduced-order modeling framework based on PINN. The reduced coefficients are predicted by a feedforward neural network trained  by minimizing the mean squared norm of the residual of the ROM.  The time derivative of the network is computed by using the automatic differentiation. This approach does not require labeled data for the training. However, the accuracy of PINN for complex nonlinear problem is often limited. To overcome this limitation, labeled data obtained from the projection of the high-fidelity snapshots used for basis generation, can be incorporated to enhance accuracy. The same approach has been used to solve inverse problems involving unknown physical quantities such as the physical viscosity \cite{Hijazi2023}. In this communication, the presence of the Forcheimer term ${\bf F_{for}}$, which cannot be directly reduced by POD, makes difficult the application of PINN coupled with ROMs for solving problems in porous media.

Furthermore, one of the drawbacks of the POD ROM Galerkin method is its lack of accuracy and its instability due to the truncation of POD modes. While negligible in terms of energy, the truncated modes play a crucial role in their interaction with the retained modes. To address this issue, numerous closure models exist, and a summary can be found in \cite{Ahmed2021}. Very recently,  a purely data-driven approach to compute the parameters of the closure
models is proposed \cite{Ahmed2023}. The influence of the truncated scales is approximated using only the information in the resolved scales with a deep neural network. More specifically, it is the projection of the terms of derivations associated with unresolved modes that is modeled. They utilize the physics-guided machine learning paradigm \cite{Pawar2021, Pawar2021b} to incorporate known physical arguments and constraints into the learning process. The approach is also coupled with Nonlinear POD to reduce the number of resolved modes to be considered. In this work, the closure model will also be based on neural network modeling. 

Finally, in many applications, the terms related to the pressure are neglected. However, incorporating these terms helps to stabilizate and to increase accuracy of reduced models \cite{Tallet2015}\cite{Leblond2011} . In this study, the pressure terms will not be directly integrated into the ROM but will be, also, modeled by a neural network.\\

To summarize, in this paper, we propose to construct a reduced model using POD-ROM Galerkin to solve flow problems in porous media. In this model, the Forchheimer term, the influence of unresolved modes, and pressure variations will be modeled using a neural network. The reduced model can be expressed as follows:
$$\fract{d{\bf a}}{dt}={\bf F}({\bf a})+{\bf R^{for}}({\bf a},\boldsymbol{\gamma})+{\bf R^{press}}({\bf a},\boldsymbol{\gamma})+{\bf R^{unres}}({\bf a},\boldsymbol{\gamma})$$
where ${\bf a}(t,\boldsymbol{\gamma})$ represents the coefficients of the POD model, depending on time $t$ and parameters $\boldsymbol{\gamma}$. ${\bf F}$ includes contributions of convection and diffusion terms. Instead of separately modeling the Forchheimer term ${\bf R^{for}}$, the pressure variations term ${\bf R^{press}}$, and the influence of unresolved modes ${\bf R^{unres}}$, these terms will be combined into a single residual $\bf{R}$, which will be modeled using a neural network.\\

This paper is organised as follows. First, the equations that describe the flow past and through a porous obstacle are presented. The reduced order model is then described and the computation of residuals with the multilayer perceptron method is detailed. The high fidelity computations are validated by comparison with results found in the literature, and the low order dynamical system enriched with the multilayer perceptron method is applied to several cases: one case for residuals obtained for one Reynolds number, a case for residuals obtained for three Reynolds numbers, and another case where the Reynolds numbers are not used to build the POD basis and the multilayer perceptron model. Finally, conclusions are drawn in the last section.

\section{Fluid flow past a porous square obstacle}
\subsection{Problem settings}
\begin{figure}[hbtp!]
	\centering
	\includegraphics[scale=0.3]{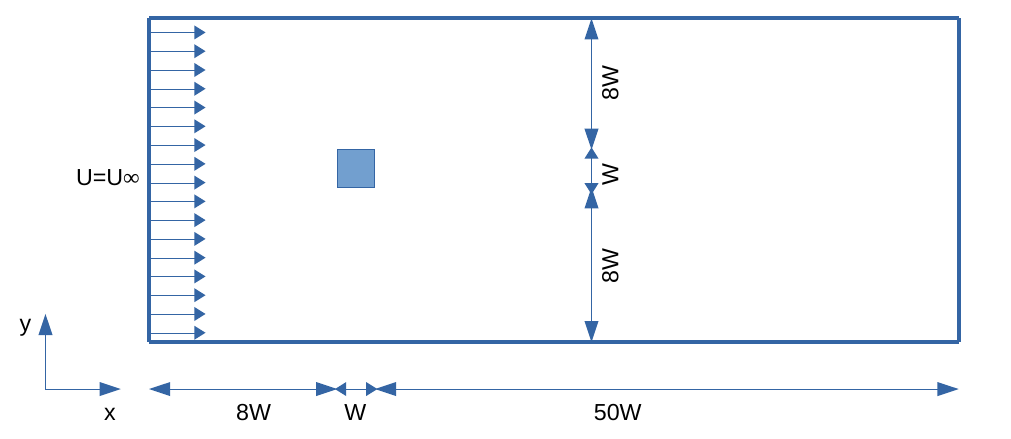} 
	\caption{Flow in an unconfined channel past a porous square obstacle}
	\label{schema_cas_etude}
\end{figure}

This study focuses on two-dimensional flow past a porous square obstacle in an unconfined channel (see Figure \ref{schema_cas_etude}). The channel boundaries, which are sufficiently far from the obstacle, do not influence the flow in the channel. The flow is laminar and incompressible. The porous medium is isotropic and homogeneous.
The conservation equations in the fluid medium and in the porous obstacle are:
\begin{itemize}
	\item[-] Continuity equation:\\
	\begin{equation}
	\nabla \cdot \textbf{u}=0
	\label{continuity}
	\end{equation}
	\item[-] Momentum conservation equation:\\
	\begin{equation}
	\fract{\partial \textbf{u}}{\partial t}+(\textbf{u}\cdot \nabla)\textbf{u}= -\fract{\nabla p}{\rho}+ \nu\Delta \textbf{u} -\bigg(\fract{\nu}{\kappa}\textbf{u} + \fract{ \tilde{F}}{\sqrt{\kappa}} \| \mathbf{u} \| \textbf{u}\bigg)\chi
	\label{momentum}
	\end{equation}
\end{itemize}
where $ \| \mathbf{u} \|=\sqrt{u^2+v^2}$, $u$ and $v$ are the velocity components according to x and y directions, $p$ is the pressure, $\rho$ is the fluid density and $\nu$ is the kinematic viscosity of the fluid. $\chi$ is a mask function that is null in the fluid region, and is equal to one in the porous region. In the momentum conservation equation of the porous material ($\chi=1$), two terms are added. The first term, called Darcy term, takes into account the viscous resistance, and it depends on the dynamic viscosity $\nu$ and the permeability $\kappa$. The second term, called Forchheimer term, is related to the inertial resistance. \\

For the porous medium which is a packed bed of spherical particles, the porosity $\porosity$ and the permeability $\kappa$ are linked with the Carman-Kozeny relationship:
\begin{equation}
Da = \fract{1}{180}\fract{\porosity^2}{\left(1-\porosity\right)^2}\left(\fract{d_p}{W}\right)^2 
\end{equation}
where $Da$ is the Darcy number $(Da=\kappa/W^2)$, 
$d_p=0.01$ is the particle diameter and $W$ is the size of the square\footnote{In this work $W=1$.}. Moreover, the inertial factor reads:
\begin{equation}
\tilde{F} = \fract{1.75}{\displaystyle{\sqrt{150}}}\porosity^{-3/2}
\end{equation}
\\

The boundary conditions are as follows : 
\begin{itemize}
	\item[-] at the inlet, a uniform velocity according to $\textbf{x}$ direction, and a zero-gradient boundary condition for the pressure;
	\item[-] at the outlet, a zero-gradient boundary condition is imposed for the velocity field, and the pressure is equal to zero, i.e. $p=0$ and $\nabla\textbf{u}=0$; 
	\item[-] symmetry boundary conditions are applied at the lower and upper boundaries of the computational domain.
\end{itemize}

\subsection{POD Galerkin reduced order model}
The aim of this study is to build a low order dynamical model based on proper orthogonal decomposition to solve the flow in the porous medium. This paragraph is dedicated to the description of the POD Galerkin reduced order model (ROM) approach.
\subsubsection{POD basis}
The proper Orthogonal Decomposition method (POD) consists in finding a reduced basis $\ensemblebase=\{\base{1}, ...,\base{q}\}$ such that the physical variables\footnote{In this study, $\fonc$ will correspond to the velocity ${\bf u}$.} $\fonc$, characteristic of the studied phenomenon, can be written according to:
\begin{equation}
\fonc(\textbf{x},\param) \approx \fonc_q(\textbf{x},\param)=\sum_{i=1}^{q}a_i(\param)\base{i}(\varspace)
\end{equation}
where $\varspace$ corresponds to the spatial variables associated to the domain $\Omega \in \mathbb{R}^2$ and $\param$ is the time or/and a physical parameter such as the Reynolds number or the permeability $\kappa$. 
\\

\noindent Consider the set $\ensemble$ composed of the function $\fonc$  taken at different instances $\gamma_i$, such that : 
$$\ensemble=\{\fonc(\varspace,\param_1)\ldots\fonc(\varspace,\param_{\nbcliche})\}$$ 
Functions $\fonc(\param_i)$  belong to  $\esp$. The POD method is a statistical technique that allows to approximate the set $\ensemble$, which consists of a large number of random datas $\{\fonc(\varspace,\param_i)\}_{i=1}^{\nbcliche}$, by a low-dimensional basis $\ensemblebase$ composed of deterministic functions $\{\base{j}(\varspace)\}_{j=1}^{\taillebase}$. It aims at finding functions $\base{j}(\varspace)$ that optimally approximate the realizations  $\{\fonc(\varspace,\param_i)\}_{i=1}^{\nbcliche}$ on average. Thus, the functions $\base{l}$ are solutions of the following maximization problem:
\begin{equation}
\label{pb_max}
\dis{\max_{\base{l}\in \esp}}\moyth{\scalth{\fonc}{\base{l}}^2} \hspace{0.1cm} \textnormal{ with } \hspace{0.1cm}
\scalth{\base{i}}{\base{l}}=\delta_{il} \hspace{0.1cm}\textnormal{
	for } 1\leq i\leq l
\end{equation}
where $\scalth{\bullet}{\bullet}$ corresponds to the inner product of
$\esp$ and $\moyth{\bullet}$ to the average operator.  Using the calculus of variations, it can be shown that the maximization problem \ref{pb_max} is equivalent to solving the following eigenvalue problem:
\begin{equation}
\int_{\Omega}R(\textbf{x},\textbf{x'})\varphi(\textbf{x})d\textbf{x'}=\lambda \varphi(\textbf{x})
\end{equation}
where $R$ is the spatial correlation tensor: $R(\textbf{x,\textbf{x'}})=<\fonc(\textbf{x},\param)\fonc(\textbf{x'},\param)>$.
The coefficients $a_i(\param)$ are calculated as:
\begin{equation}
a_i=(\fonc(\textbf{x},\param),\base{i}(\textbf{x}))
\label{aiPOD}
\end{equation}
\\
In practice, for numerical simulations, the evaluation of the tensor $R$ is a very large computational task. Sirovich \cite{Sirovich87} showed that the eigenfunction can be expressed in terms of the snapshots $\{\fonc(\textbf{x},\boldsymbol{\gamma_i})\}_{i=1}^{\nbcliche}$: 
\begin{equation}
\base{n}(\textbf{x})=\sum_{k=1}^{\nbcliche}a_n(\boldsymbol{\gamma_k})\fonc(\textbf{x},\boldsymbol{\gamma_k})
\end{equation}
In this expression, the coefficients $a_n(\boldsymbol{\gamma_k})$ must be determined. To do that, the following matrix eigenvalue problem must be solved:
\begin{equation}
\textbf{C} \textbf{a}=\lambda \textbf{a}
\label{eigenvalue}
\end{equation}
where $\nbcliche$ is the number of snapshots and $\textbf{C}$ is the correlation tensor defined by:
\begin{equation}
C_{ij}=\fract{1}{\nbcliche}\int_{\Omega}w(\textbf{x},\boldsymbol{\gamma_i})w(\textbf{x},\boldsymbol{\gamma_j})d\textbf{x}
\label{correlation_tensor}
\end{equation}
\\

\noindent Since the POD basis is optimal in an energetic sense, a small number of modes $q$ contain most of the energy and it is possible to approximate the solution as follows : $\displaystyle \fonc(\textbf{x},\param)\approx\sum_{i=1}^q a_i(\param)\varphi_i(\textbf{x})$. Once the POD reduced basis is obtained, to calculate the evolution of $\fonc$ as a function of $\param$, a system of equations depending on $a_i(\param)$ must be built. This system is obtained by performing a Galerkin projection of the equations governing the problem onto the most energetic modes. This system has a small number of differential equations and unknowns.\\ 
\subsubsection{Reduced order model for flows in porous media}
To obtain the reduced order model (ROM), the first step consists in splitting the velocity and pressure fields into a mean and a fluctuating part:
\begin{equation}
\label{equ:fluct}
\left\{
\begin{aligned}
\textbf{u}(\textbf{x},t,\paramB)&=\overline{\textbf{u}}(\textbf{x})+\textbf{u}'(\textbf{x},t,\paramB)\\
p(\textbf{x},t,\paramB)&=\overline{p}(\textbf{x})+p'(\textbf{x},t,\paramB)
\end{aligned}
\right.
\end{equation}
where $\paramB$ corresponds to physical parameters such as the Reynolds number or the permeability $\kappa$. 
The POD decomposition is applied to the fluctuating part of the velocity fields:
\begin{equation}
\label{equ:decomp}
\textbf{u}'(\textbf{x},t,\paramB)={\bf u'_{res}}(\textbf{x},t,\paramB)+{\bf u'_{unres}}(\textbf{x},t,\paramB)=  \sum_{i=1}^{N_u}a_i(t,\paramB) \basevit{i}(\textbf{x})+\sum_{i=Nu+1}^{Ns}a_i(t,\paramB) \basevit{i}(\textbf{x})
\end{equation}
where ${\bf u'_{res}}$ corresponds to the contribution of the $N_u$ most energetic modes that will be kept, and ${\bf u'_{unres}}$ to the contribution of the less energetic modes that will be omitted. While negligible in terms of energy, the truncated modes play a crucial role in their interaction with the retained modes and must be modelled. A summary of the closure  model can be found in \cite{Ahmed2021}.

Injecting the expressions \ref{equ:fluct} and \ref{equ:decomp} into the Navier Stokes equations and performing a Galerkin projection  onto the first velocity modes $\basevit{i}$ with $i=1,\ldots,N_u$ yields the following differential equation:
\begin{equation}
\fract{da_i}{dt}=\sum_{j=1}^{N_u}\sum_{k=1}^{N_u}C_{ijk}a_j a_k+\sum_{j=1}^{N_u}(B_{ij}+D_{ij})a_j+E_i+R^{for}_i+R^{unres}_i+R^{press}_i
\label{ODE_an}
\end{equation}
where:
\begin{equation}
\begin{aligned}
B_{ij}&=(-\overline{\textbf{u}}\cdot \nabla \basevit{j}-\boldsymbol{\varphi_j^u}\cdot \nabla \overline{\boldsymbol{u}},\basevit{i})\\
D_{ij}&=(\nu \Delta \basevit{j}-\fract{\nu}{\kappa}\chi \basevit{j},\basevit{i})\\
C_{ijk}&=(\basevit{j} \cdot \nabla \basevit{k},\basevit{i})\\
E_i&=(-\overline{\textbf{u}}\cdot \nabla \overline{\textbf{u}}+\nu \Delta \overline{\boldsymbol{u}} -\fract{1}{\rho}\nabla\overline{p}-\fract{\nu}{\kappa}\chi\overline{\boldsymbol{u}},\basevit{i})\\
\end{aligned} 
\end{equation}
$R^{unres}_i$ corresponds to the contribution of the unresolved scales projected onto the POD mode $\basevit{i}$ and $R^{press}_i= -\fract{1}{\rho}(\nabla p',\basevit{i})$ is the contribution of the fluctuating pressure. The part dedicated to the Forchheimer term is given by:
$$R^{for}_i= -\fract{ \tilde{F}}{\sqrt{\kappa}}(\chi\|{\bf u_{res}}\|\sum_{j=1}^{N_u}a_j \basevit{j}\,\basevit{i})$$
where ${\bf u_{res}}({\bf x})$ is the norm of the velocity for each point ${\bf x}$. This norm cannot be expressed in reduced form.  At each new time step $R^{for}_i$ needs to be recalculated by projection, which significantly increases the computational cost, rendering the reduced model inefficient. 
To model the influence of these different terms, we propose to group them into a single term $R_i=R^{for}_i+R^{press}_i+R^{unres}_i$ and model it using a neural network.
Thus, the reduced model to be solved can be written as follows:
\begin{equation}
\label{ROM}
\fract{d{\bf a}}{dt}={\bf F}({\bf a},\boldsymbol{\gamma})+{\bf R}({\bf a},\boldsymbol{\gamma})
\end{equation}
where  ${\bf F}$ includes contributions of convection and diffusion terms. ${\bf R}$ is the unknown residual depending on temporal POD coefficients ${\bf a}$ and parameters $\boldsymbol{\gamma}$ which will be modeled using a multilayer perceptron approach.

The construction of the neural network is done by using the coefficients ${\bf a_{POD}}$ obtained from the POD. Then, the training datas for ${\bf R}$ are estimated as follows:
\begin{equation}
\label{build_MLP}
{\bf R}({\bf a_{POD}},\boldsymbol{\gamma})\simeq \fract{d{\bf a_{POD}}}{dt}-{\bf F}({\bf a_{POD}},\boldsymbol{\gamma})
\end{equation}
For the calculation, the temporal terms $\displaystyle \fract{d {\bf a_{POD}}}{dt}$ were evaluated with a third order finite difference.
\\

\noindent In this work, the ROM, defined by equation \ref{ROM}, was solved either without residuals or with residuals computed with multilayer perceptrons. A Runge Kutta second order method was used to solve (\ref{ODE_an}) with or without residuals. 

\section{Computation of the residuals with multilayer perceptrons}
This study aims at taking into account the residuals ${\bf R}$ to improve the results of the low order dynamical system. To do that, the computed snapshots provided a data set associating the exact temporal coefficients ${ \bf a}$ with the residuals ${\bf R}$, and supervised machine learning algorithms were then employed.

\subsection{Multilayer perceptrons}
To solve that problem, this study aims at learning a hypothesis function that gives the best mapping of the temporal coefficients ${\bf a}=\hspace*{0.05cm}^T(a_1(t),\cdots,a_{N_u}(t))$ to the residuals ${\bf R}= \hspace*{0.05cm}^T(R_1(t),\cdots,R_n(t))$ (where $N_u$ is the number of kept modes) from the knowledge of $N_s$ samples which associate values of ${\bf a}$ to values of ${\bf R}$. These samples are obtained from the snapshots calculated during the sampling time. Among supervised machine learning algorithms, multilayer perceptrons (MLP), which are capable of learning non linear problems, were considered. However, it is to be noticed that the Random Forest method, the Support Vector Machine method, the Gradient Boosting method were tested for this study, but they did not provide better results than the multilayer perceptron method. We search a hypothesis function ${\bf l}$ :
\begin{equation}
{\bf l}({\bf a})\approx {\bf R }
\label{MLP}
\end{equation}
In the remainder of this paper, we consider that we have a dataset either of residuals ${\bf R}\in \mathbb{R}^{N_s\times N_u}$ associated with temporal coefficients ${\bf A}\in \mathbb{R}^{N_s\times N_u}$ or residuals ${\bf R}$ associated with temporal coefficients and certain values of the Reynolds number ${\bf A}\in \mathbb{R}^{N_s\times (N_u+1)}$.\\

\noindent Multilayer perceptrons are feed forward neural networks with fully connected hidden layers \cite{geron2019hands}, \cite{haykin1999neural}. In this paper, the Keras and Scikit-Learn libraries are used to solve the problem (\ref{MLP}). In particular, Keras was considered for building, training, evaluating and running the MLP, whereas Scikit-Learn was employed to perform train-test splits and data scaling. Apart from the number of neurons and hidden layers, the following choices were made regarding the other important configuration parameters of an MLP model with Keras:
\begin{itemize}
	\item[-] the activation function considered for the hidden layers was the rectified linear activation (ReLU);
	\item[-] the Adam optimizer, initialized to its default values, was used when compiling Keras models;
	\item[-] the loss function to be minimized during the training was the mean squared error (MSE), defined as
	\begin{equation}
	\text{MSE}({\bf R}, {\bf \hat{R}}) = \dfrac{1}{N_s} \sum_{i=1}^{N_s} \sum_{j=1}^{N_u} \left( R_i^j - \hat{R}_i^j \right)^2
	\end{equation}
	where $R_i^j$ and $\hat{R}_i^j$ are the true and predicted residuals, respectively, for mode $j$ and sample $i$.
\end{itemize}
Note that the seeds used to initialize random number generators were fixed when splitting the dataset in order to have reproductible results at most, being given the stochastic nature of machine learning algorithms. 
\subsection{Cross-validation}

Since the only way to know how well a model will generalize to new data is to try it on new data, a first approach to train and validate a model is 
to split the data into two sets, namely the training and test sets, where the model learns from the training data and is tested on the test data. But, as we sought to tune our model to find one that performed well on new data, this approach would lead to overfitting the model and hyperparameters for this particular test set and was unlikely to yield a model that performed as well on new data. A better approach is to introduce a validation set dedicated to select the model that performs best.   The latter is then used to evaluate the error on the test set.  Although this approach is more robust than the previous one, it leads to complications related to the choice of the size of the validation set.

A solution to find the ideal values of the hyperparameters of the MLP, such as the number of layers and the number of neurons, is to use the well-known cross-validation technique. 
It consists in performing repeated cross-validations, using many small validation sets. 
Each model is evaluated once per validation set, after being trained on the rest of the data. 
A much more accurate measure of the model performance can be then obtained by averaging out all model evaluations over the validation sets.

In this paper, the \textit{RepeatedKFold} method of Scikit-Learn was used to generate the folds. 
It depends on the parameters $n_{\text{folds}}$ and $\text{cv}_{\text{repeats}}$, for the number of folds and the number of repetitions of the cross-validation, respectively. 

\subsection{Features/Targets scaling}

One of the disadvantages of an MLP is that it is sensitive to feature scaling. 
A remedy is to normalize each column of the features by scaling each feature to lie between two values or to resemble standard normally distributed data. 
The former method was considered in this work. Moreover, as the targets $R_i^j$ were very small, a min-max scaling in $[0, 1]$ was applied to them. 
Although the standardization of the targets is not necessary in general for MLP, it was mandatory in the present case due to numerical reasons. Of course, special care was taken when scaling features and targets in order to avoid data leakage: 
scalers were computed on the training set only before being applied to training, validation, or test data.

\subsection{Early stopping}

In machine learning, early stopping can be seen as a form of regularization tool by stopping the training when validation errors begin to degrade, which in particular helps to avoid overfitting. This technique was considered in this study in the following way. The MSE on the validation sets was monitored and the training was stopped when the number of epochs without improvement of the validation loss was equal to the parameter $p$. Finally, the model weights, $p$ epochs backward, were restored and saved if necessary.

\subsection{Algorithm}\label{subsec:algorithm}

Finally, being given two matrices ${\bf A}, {\bf R} \in \mathbb{R}^{N_s \times N_u}$, 
the procedure developed to solve equation \eqref{MLP} by building and validating a MLP model that can predict residuals from temporal coefficients is as follows.

\begin{enumerate}
	\item Define (i) the proportion of the data used to test the model on new data, $t_{\text{size}} = 0.1$, (ii) the maximum number of epochs as long as the learning process is not interrupted by the early stopping, $n_{\text{epochs}} = 100$, (iii) the number of epochs before the early stopping is triggered, $n_{\text{patience}}=3$, 
	(iv) the number of folds in the cross-validation, $n_{\text{folds}} = 9$, (v) the number of repetitions of the cross-validation, $\text{cv}_{\text{repeats}}=3$, and (vi) the number of repetitions when computing the last model, $n_{\text{repeats}}=5$ (to deal with the stochastic nature of machine learning algorithms.).
	\item Define the search for hyperparameters:
	\begin{itemize}
		\item[-] features scaling method : $\text{min-max}([0, 1])$ 
		\item[-] number of layers of the neural network, $n_{\text{layers}}$, and their numbers of neurons, $n_{\text{neurons}}$;
	\end{itemize}
	\item Split matrices ${\bf A}$ and $\bf{R}$ into training sets $\left( {\bf A_{\text{train}}}, {\bf R_{\text{train}}} \right)$ and test sets $\left( {\bf A_{\text{test}}}, {\bf R_{\text{test}}} \right)$, according to $t_{\text{size}}$;
	\item Perform the cross-validation with $\left( {\bf A_{\text{train}}}, {\bf R_{\text{train}}} \right)$ according to $n_{\text{folds}}$ and $\text{cv}_{\text{repeats}}$, 
	for all potential models defined by the set of hyperparameters. Each of the validation sets is considered when monitoring the MSE for the early stopping;
	\item Select hyperparameters according to the model with the lowest MSE on average over the validation sets;
	\item Split matrices $\left( {\bf A_{\text{train}}},{\bf R_{\text{train}}} \right)$ into $\left( {\bf \tilde{A}_{\text{train}}}, {\bf \tilde{R}_{\text{train}}} \right)$ and $\left( {\bf A_{\text{val}}}, {\bf R_{\text{val}}} \right)$ with the same number of samples in validation sets $\left( {\bf A_{\text{val}}}, {\bf R_{\text{val}}} \right)$ as in test sets $\left( {\bf A_{\text{test}}}, {\bf R_{\text{test}}} \right)$;
	\item Train $n_{\text{repeats}}$ models with these hyperparameters using $\left( {\bf \tilde{A}_{\text{train}}}, {\bf \tilde{R}_{\text{train}}} \right)$ and $\left( {\bf A_{\text{val}}}, {\bf R_{\text{val}}} \right)$. 
	Select the one with the smallest MSE on the validation set;
	\item evaluate the performance of the selected model on the test set ${\bf R_{\text{test}}}$ using ${\bf A_{\text{test}}}$.
\end{enumerate}
Note that with such values for $t_{\text{size}}$ and $n_{\text{folds}}$, equal to $0.1$ and $9$, the size of the test set and validation sets in the cross-validation are the same.

\section{Results and discussion}
\subsection{Validation of the high fidelity computations}
\noindent In this section, the CFD model used to solve the flow around and in the porous obstacle is first validated. The high fidelity computations were validated by comparing the drag coefficient, the lift coefficient and the Strouhal number, for three different cases: ($Da=10^{-6}$, $Re=100$), ($Da=10^{-6}$, $Re=50$) and ($Da=10^{-3}$, $Re=10$). $Re$ is the Reynolds number and $Da$ the Darcy number. The density and the kinematic viscosity were, respectively, $\rho=1kg/m^3$ and $\nu=1.506\times10^{-5}m^2/s$.

\begin{figure}[hbtp!]
	\centering
	\includegraphics[scale=0.5]{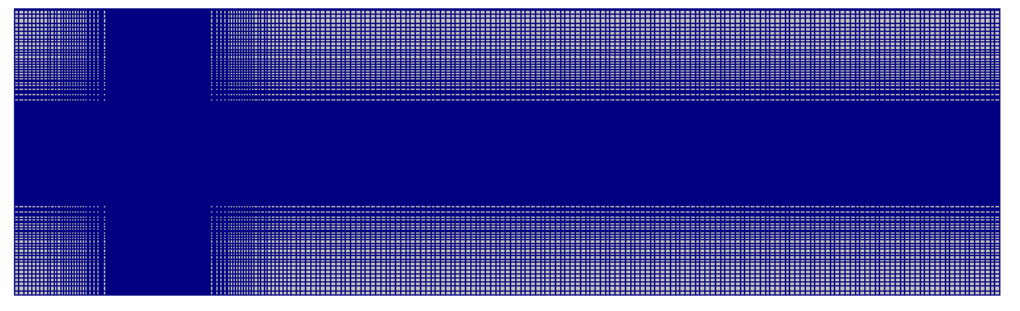}$  $
	\caption{Non uniform grid}
	\label{maillage}
\end{figure}

The flow was computed with the finite volume code OpenFOAM. The PISO algorithm was used to simulate the flow. The Gauss linear scheme was used to discretize the gradient, the divergence and the laplacian schemes. The time derivatives were discretized with the Euler scheme. A structured non uniform grid of $171500$ cells, with $10000$ cells in the porous square cylinder, was built (see Figure \ref{maillage}).

In Table \ref{CdClSt} the drag coefficient, the r.m.s. lift coefficient and the Strouhal number are listed. We can notice a good agreement between the results of the present study and those obtained by Anirudh and Dhinakaran \cite{Anirudh}, Sharma and Eswaran \cite{Sharma}, Dhinakaran and Ponmozhi \cite{Dhinakaran}.\\ 
\begin{table}[hbtp!]
	\centering
	\begin{tabular}{c|c|l|l|l}
		$Re$ & $Da$ &       $\;\;\;\;\;\;\;\;\;\;\;\;\;\;\;\;\;\;\;\;\;C_D$ & $\;\;\;\;\;\;C_{L rms}$ & $\;\;\;\;\;\;\;\;\;\;\;\;\;St$\\
		\cline{3-5}
		&      & $\;\;\;a\;\;\;\;\;\;\;\;\;\; b \;\;\;\;\;\;\;\;\;\;\;c \;\;\;\;\;\;\;\;\;\;d$ & $\;\;a \;\;\;\;\;\;\;\;\; c$ & $\;\;\;a \;\;\;\;\;\;\;\;\;b \;\;\;\;\;\;\;\;\;c$\\
		\hline    
		10  & $10^{-3}$&3.296 $\;\;\;3.241\;\;\;\;\;\;\;\;\;\;\;\;\;\;\;3.261 $   &       &       \\
		50  & $10^{-6}$&1.691 $\;\;\;1.671\;\;\;1.657$     &$0.043\;\;\;\;0.05$       &$0.119\;\;\;0.117\;\;\;0.118$       \\
		100  & $10^{-6}$&1.514 $\;\;\;1.512\;\;\; 1.495$&$0.187\;\;\;\;0.192$       &$0.149\;\;\;\;0.15\;\;\;\;0.149$
	\end{tabular}
	\caption{Drag coefficient, lift coefficient and Strouhal number. (a) Present study, (b) Anirudh and Dhinakaran \cite{Anirudh}, (c) Sharma and Eswaran \cite{Sharma}, (d) Dhinakaran and Ponmozhi \cite{Dhinakaran}.}
	\label{CdClSt}
\end{table}

In the subsequent part of this work, the computations carried out with OpenFOAM were used to build a reduced order model and to apply the artificial neural networks approach to predict the velocity field in the computational domain.  

\subsection{Computations performed with the high fidelity solver}
\begin{figure}[hbtp!]
	\begin{subfigure}{0.5\textwidth}
		\centering
		\includegraphics[width=\textwidth]{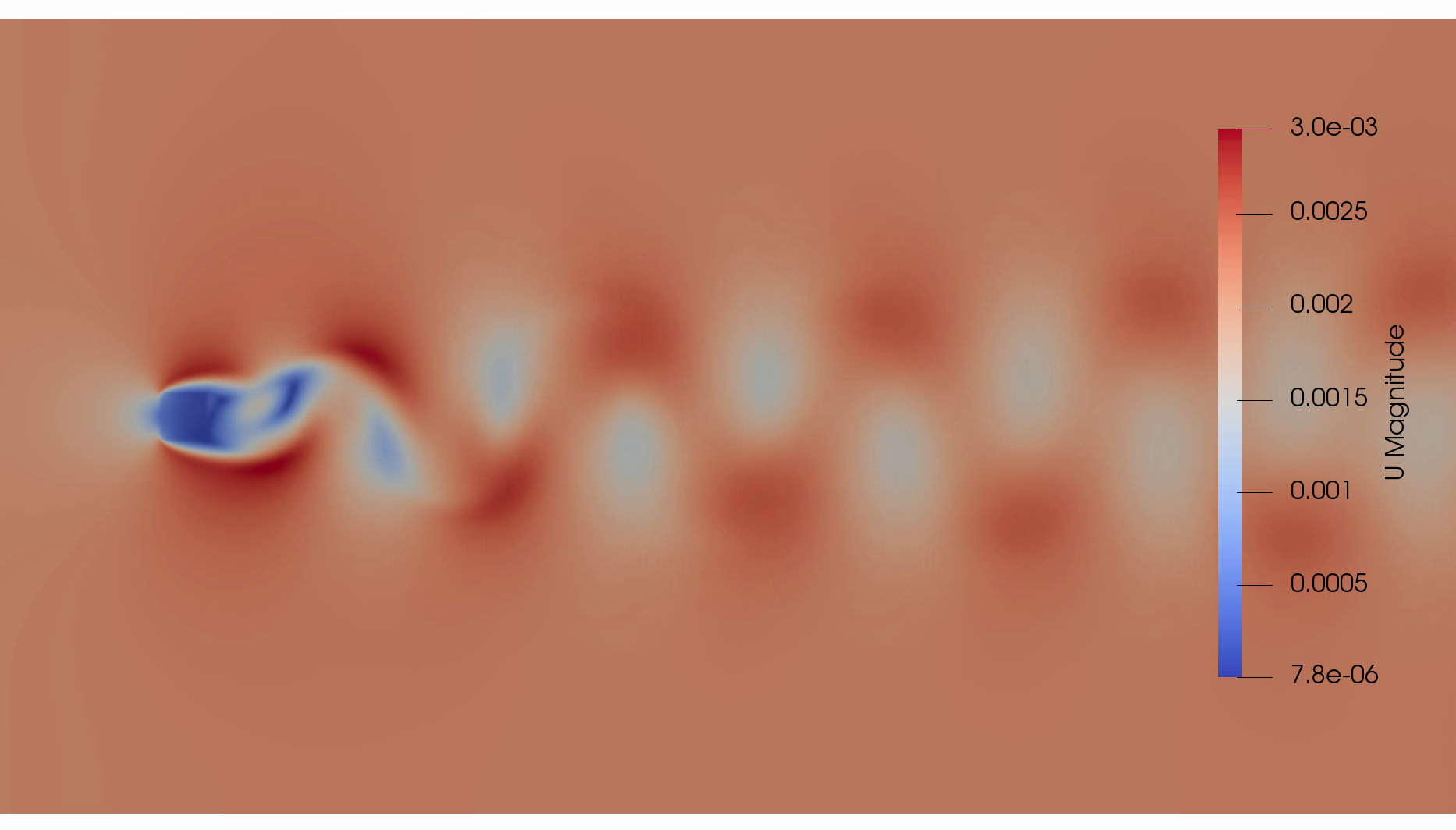}
	\end{subfigure}
	\begin{subfigure}{0.5\textwidth}
		\centering
		\includegraphics[width=\textwidth]{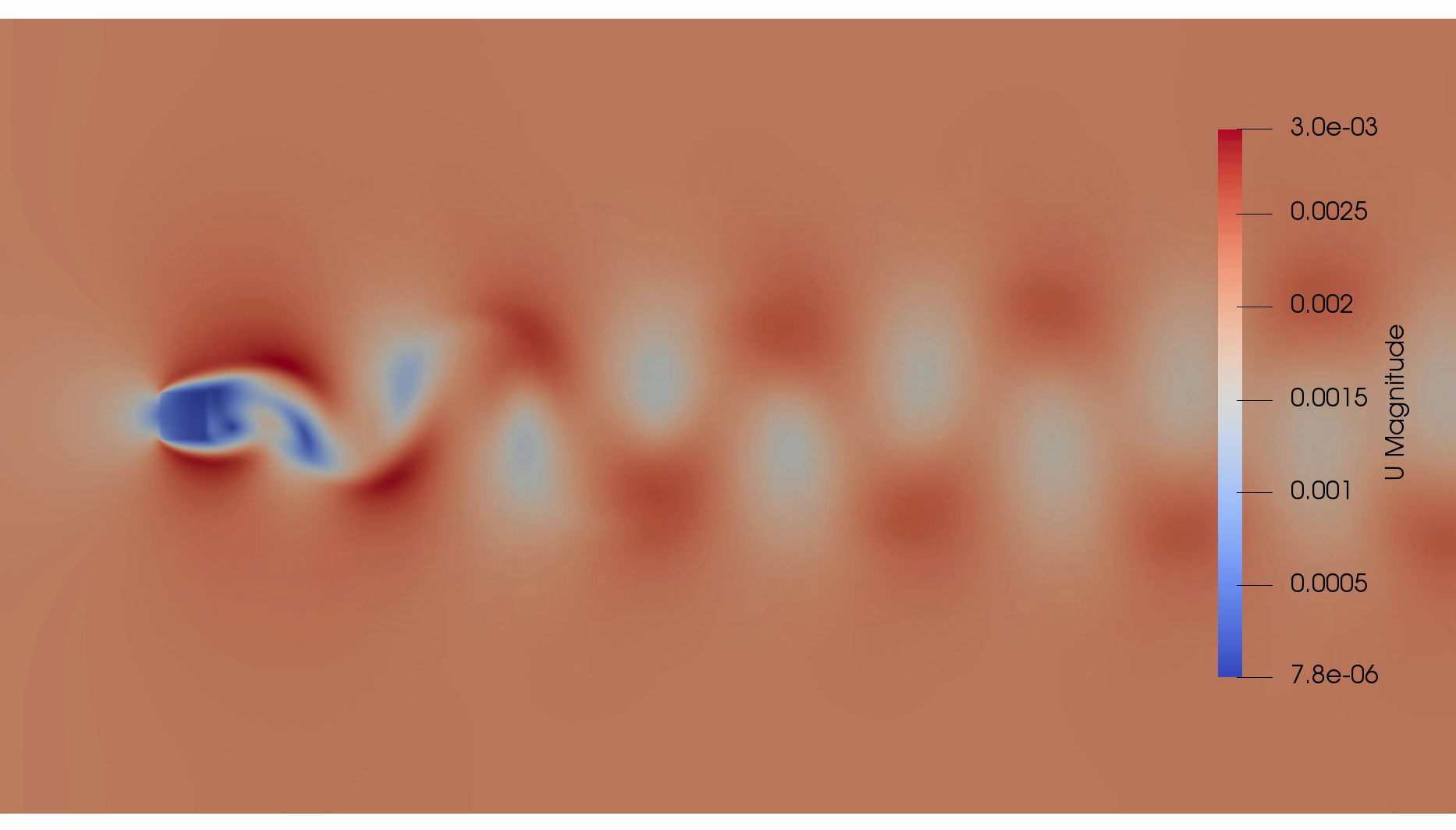}
	\end{subfigure}
	\caption{Instantaneous isovalues of the velocity magnitude for a Reynolds number of 140}
	\label{velocityRe140}
\end{figure}

Computations were performed with the Navier-Stokes equations (\ref{continuity}) and (\ref{momentum}) for a density and a kinematic viscosity equal to $\rho=1kg/m^3$ and $\nu=1.506\times 10^{-5}m^2/s$. The Darcy number was $Da=0.0007355$ and various Reynolds numbers $Re\in [110,160]$ were considered. Figure \ref{velocityRe140} depicts the instantaneous isovalues of the velocity magnitude for a Reynolds number of 140, at two successive time instants. A Von Karman vortex street can be noticed, and the velocity inside the porous obstacle is not null. Similar features were noted for the other Reynolds numbers studied in this work and they are not shown here.

\begin{figure}[hbtp!]
	\begin{tabular}{cc}
		\includegraphics[width=0.5\textwidth]{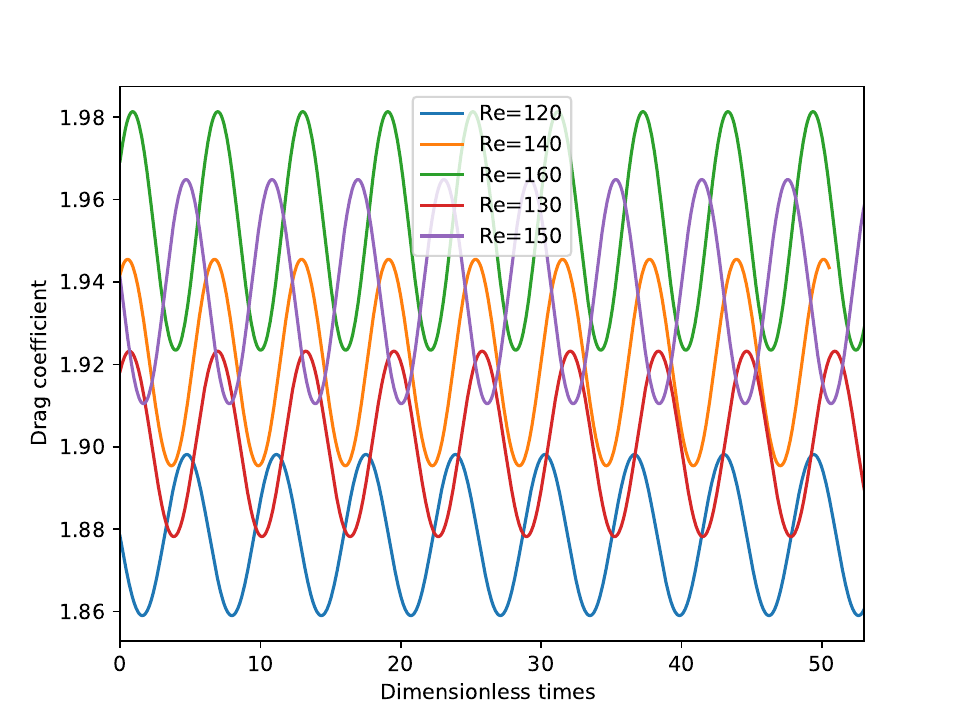}&\includegraphics[width=0.5\textwidth]{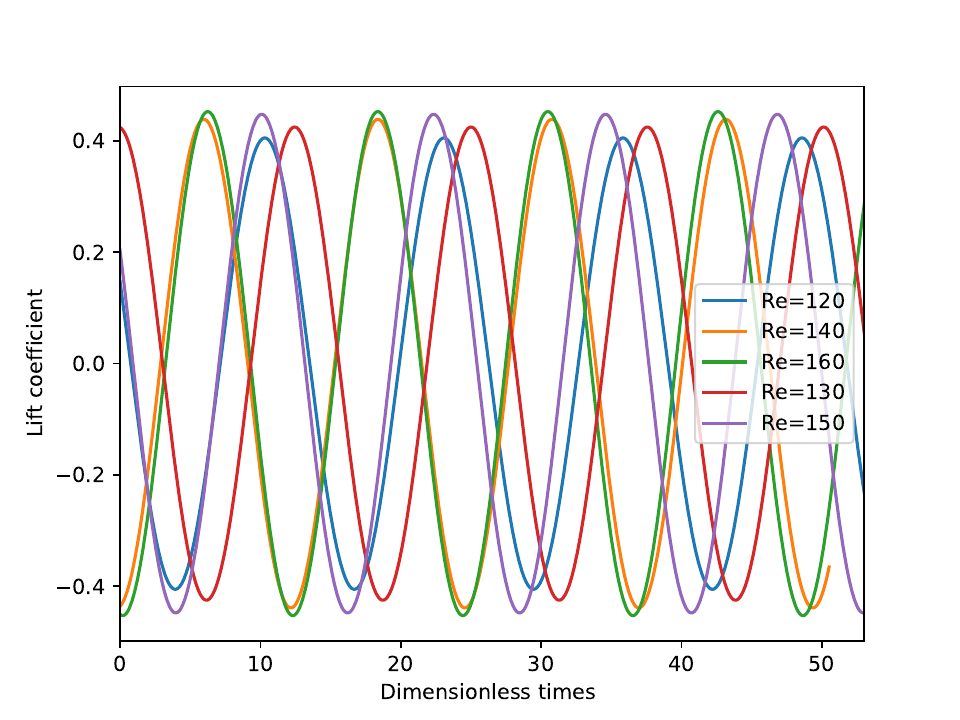}\\
		\footnotesize{a)  Drag coefficient}& \footnotesize{b)  Lift coefficient}\\[0.3cm]
	\end{tabular}
	\caption{Drag and lift coefficients obtained for all Reynolds numbers. }
	\label{fig:cdclfull}
\end{figure}

Figure \ref{fig:cdclfull} displays the drag and lift coefficients obtained for the considered Reynolds numbers and four periods of time. It can be seen that the higher the Reynolds number, the larger the drag coefficient, and that an increasing Reynolds number leads to a slight elevation of the lift coefficient.

\subsection{Results obtained with the low order reduced model without and with residuals for one Reynolds number}
\subsubsection{Application of the POD method}
\begin{figure}[hbtp!]
	\centerline{\includegraphics[width=0.75\textwidth]{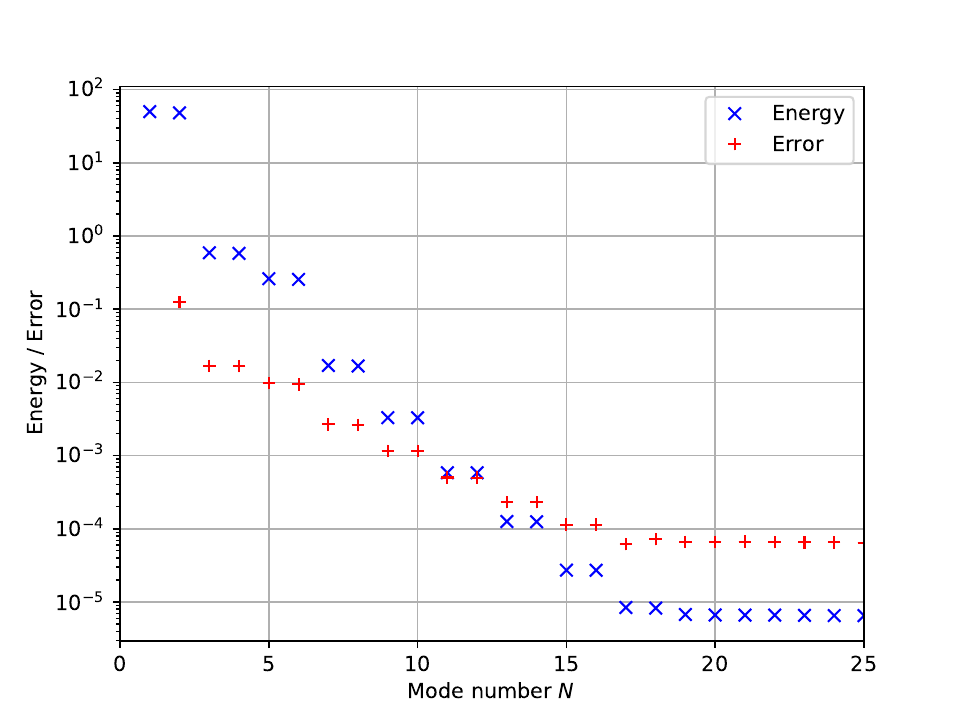}}
	\caption{Amount of fluctuating energy of each POD mode (blue curve) and maximum relative error between the solution obtained by the Navier-Stokes equations and the solution built by POD computed by keeping $N$ POD modes (red curve).}
	\label{fig:unReyn_energy}
\end{figure}

For that first case, only one Reynolds number $Re=110$ was considered. To build the POD basis, one thousand regularly spaced snapshots located in four periods of time were recorded. With the temporal correlation tensor (\ref{correlation_tensor}) obtained with these snapshots, the eigenvalue problem (\ref{eigenvalue}) was solved. In figure \ref{fig:unReyn_energy} the fluctuating energy and the maximum relative error between the solution of the Navier-Stokes equations and the velocity computed by POD was plotted. This error was obtained as follows:
\begin{equation}
\label{erreur_recPOD}
err(N)=\underset{t}{sup} \fract{\|\vitfull(t)-\vitPOD(t))\|_{L^2(\Omega)}}{\|\vitfull(t)\|_{L^2(\Omega)}}
\end{equation}
In this expression, $\vitPOD(t)$ is the velocity calculated by keeping $N$ POD modes, $\|\bullet\|_{L^2(\Omega)}=\displaystyle{\sqrt{\integ{\Omega}{}(\bullet)^2\dx}}$ is the $L^2(\Omega)$ norm and $\vitfull(t)$ is the velocity computed with the CFD model. 

In this figure, one can see that the amount of fluctuating energy and the error decrease for an increasing number of POD modes, and that 16 POD modes are sufficient to obtain an error around $10^{-4}$.  Furthermore, the error stagnates for more than 16 modes, thus 16 modes were kept.

\subsubsection{Creation of the MLP model}
\begin{figure}[!hbtp]
	\centering
	\includegraphics[width=0.75\textwidth]{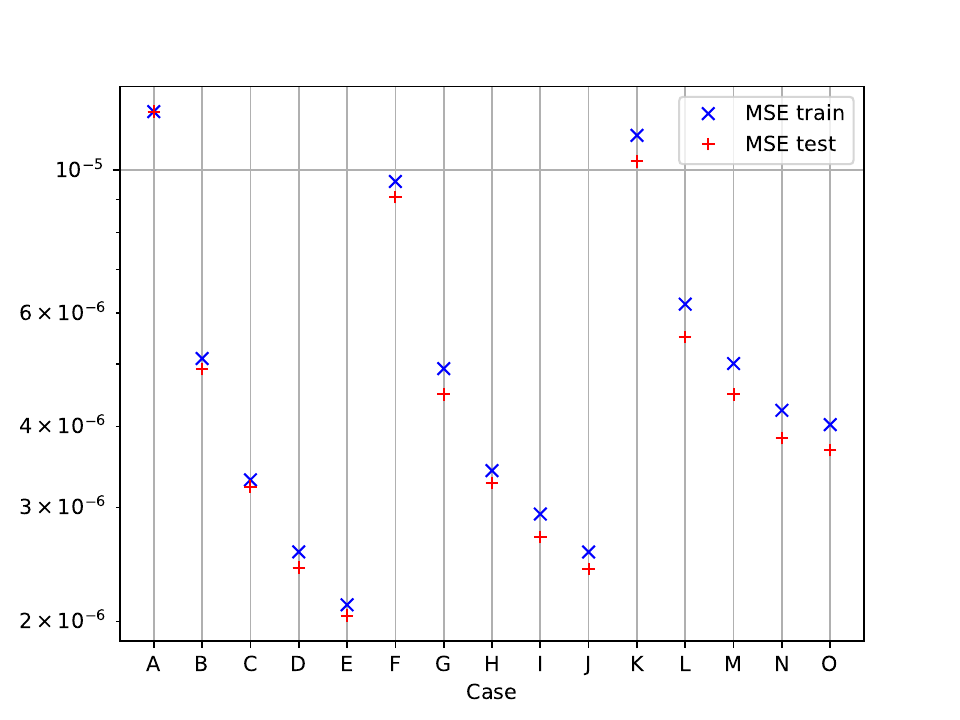}
	\caption{Mean MSE errors over the validation sets for different values of the hyperparameters in the case of 1000 data, 16 modes and a constant Reynolds number.}\label{fig:study1:losses}
\end{figure}

\begin{table}[hbtp!]
	\centering
	\begin{filecontents*}{df_all_results_pour_latex1.csv}
Case,Nblayers,Nbneurons,Nbparams,MeanMSEtrain,MeanMSEtest,Meannbepochs
A,1,200,6616,1.23e-05,1.23e-05,48
B,1,400,13216,5.10e-06,4.91e-06,39
C,1,600,19816,3.31e-06,3.23e-06,33
D,1,800,26416,2.56e-06,2.42e-06,31
E,1,1000,33016,2.12e-06,2.04e-06,26
F,2,200,46816,9.59e-06,9.07e-06,25
G,2,400,173616,4.92e-06,4.49e-06,17
H,2,600,380416,3.42e-06,3.27e-06,15
I,2,800,667216,2.93e-06,2.70e-06,13
J,2,1000,1034016,2.56e-06,2.41e-06,12
K,3,200,87016,1.13e-05,1.03e-05,16
L,3,400,334016,6.19e-06,5.51e-06,11
M,3,600,741016,5.01e-06,4.49e-06,9
N,3,800,1308016,4.24e-06,3.84e-06,8
O,3,1000,2035016,4.03e-06,3.68e-06,7
\end{filecontents*}
\csvreader[head to column names,tabular=|c|c|c|c|c|c|,
	table head=\hline \rowcolor{white!50} Case &  Nb layers&  Nb neurons &  Nb params &  MSE train &  MSE test ,
	late after head=\\\hline\rowcolor{darkgray!10},
	late after line=\csvifoddrow{\\\rowcolor{darkgray!10}}{\\\rowcolor{darkgray!25}}\hline]
	{df_all_results_pour_latex1.csv}{Case=\Case,Nblayers=\Nblayers, Nbneurons=\Nbneurons,Nbparams=\Nbparams,MeanMSEtrain=\MeanMSEtrain,MeanMSEtest=\MeanMSEtest}
	{\Case &\Nblayers & \Nbneurons  & \Nbparams  & \MeanMSEtrain  &  \MeanMSEtest  }
	\caption{Mean Square Error on the train set (MSE train) and on the test set (MSE test) with respect to the number of layers and the number of neurons.}
	\label{table:unRey_msecase}
\end{table}
To calculate the residuals of the low order dynamical system with the multilayer perceptron model, the algorithm presented in paragraph (\ref{subsec:algorithm}) was applied to the 1000 snapshots and 16 modes. As a first step, we considered $n_{\text{layers}} \in \left\{ 1, 2, 3 \right\}$ and $n_{\text{neurons}} \in \left\{ 200, 400, 600, 800, 1000 \right\}$.  This results in 15 cases for the hyperparameter search which are depicted in table \ref{table:unRey_msecase} and in figure \ref{fig:study1:losses}.  It is observed that for a given number of layers, the MSE error on the train and test sets decreased for an increasing number of neurons. In addition, the best results were obtained for the 1-layer networks. It can also be highlighted that the higher the number of layers and neurons, the higher the number of parameters. For the 1-layer networks, we computed the performance of the reduced order models during the sampling period. 
\\

In table \ref{table:unReyn_ROMcase}, $\varepsilon_a$ is the errror calculated with the temporal coefficients according to:
$$\varepsilon_a=\fract{\|{\bf a_{POD}}\|_{F}-\|{\bf a_{case}}\|_{F}}{{\bf\|a_{POD}}\|_{F}}$$
where $\|{\bf a}\|_{F}$ is the Froebenius norm of matrix ${\bf a}$ where the temporal coefficients of each mode obtained for all snapshots in the sampling interval were stored. $\bf a_{POD}$ is the matrix that contains the reference coefficients and $\bf a_{case}$ is the matrix that includes the coefficients obtained with the reduced order model without or with residuals. The error relative to the lift coefficient is as follows :
$$\varepsilon_L=\bigg{\|}\fract{{\bf C_{L_{pod}}}-{\bf C_{L_{case}}}}{{\bf C_{L_{pod}}}}\bigg{\|}$$
where $\|{\bf C_L} \|$ is the euclidian norm of vector ${\bf C_L}$ which includes the lift coefficients predicted for all snapshots in the sampling interval. The error $\varepsilon_d$ related to the drag coefficient is calculated in a similar manner. CPU is the CPU time required to solve the reduced order models during the sampling period. 
\begin{table}[hbtp!]
	\centering
	\begin{filecontents*}{Erreur_ROM_nbneuronnes1.csv}
Case,Nbneurons,NormeFroebat,NormeCL,NormeCd,CPUTimes
ROM,/,1.94e+00,3.34e-01,9.48e-03,5
A,200,2.54e-03,3.03e-04,5.95e-06,37
B,400,2.62e-03,3.91e-04,6.15e-06,37
C,600,2.16e-03,1.52e-04,6.08e-06,38
D,800,1.63e-03,8.12e-05,4.79e-06,38
E,1000,2.16e-03,1.74e-04,4.76e-06,39
\end{filecontents*}
\csvreader[head to column names,tabular=|c|c|c|c|c|c|,
	table head=\hline \rowcolor{white!50} Case &Nb neurons &  $a_t$ error ($\varepsilon_a$)&  $C_L$ error  ($\varepsilon_L$) &  $C_d$ error ($\varepsilon_d$)&  CPU ,
	late after head=\\\hline\rowcolor{darkgray!10},
	late after line=\csvifoddrow{\\\rowcolor{darkgray!10}}{\\\rowcolor{darkgray!25}}\hline]
	{Erreur_ROM_nbneuronnes1.csv}{Case=\Case, Nbneurons=\Nbneurons,NormeFroebat=\NormeFroebat, NormeCL=\NormeCL,NormeCd=\NormeCd,CPUTimes=\CPUTimes}
	{\Case & \Nbneurons&\NormeFroebat & \NormeCL  & \NormeCd & \CPUTimes \ s}
	\caption{Influence of the MLP model on the results obtained with the reduced order model.}
	\label{table:unReyn_ROMcase}
\end{table}
In this table, we can first notice that the errors produced by the ROM (reduced order models in which the residuals were omitted) are much higher than those obtained with the reduced order models where the residuals calculated with MLP were included. For this case, the CPU time was very small. The ROM MLP with 800 neurons led to the smallest errors ($0.00812\%$ for the lift coefficient and $0.000479\%$ for the drag coefficient). For an increasing number of neurons, the CPU time slightly increases (between 37 and 39s), but it is nevertheless negligible in comparison with the CPU time needed to solve the Navier-Stokes equations (2 hours and 40 minutes). Let us also mention here that the CPU time needed to construct the MLP model was 309s.\\ 

Once the hyperparameter search was complete, a model with 1 hidden layer and 800 neurons was selected.
\begin{figure}[hbtp!]
	\centerline{\includegraphics[width=0.75\textwidth]{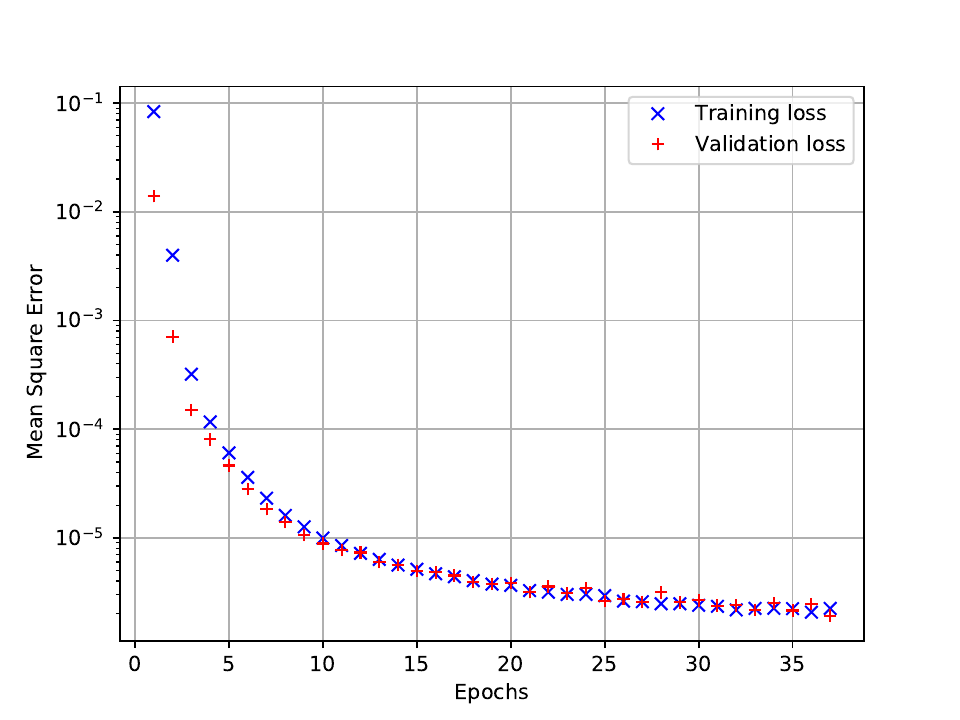}}
	\caption{Training and validation losses with respect to the number of epochs for Re=110.
		The MSE is presented in logarithmic scales.}
	\label{fig:unReyn_loss}
\end{figure}
Figure \ref{fig:unReyn_loss} shows the evolution of training and validation losses with respect to the number of epochs. 
It can be noticed that the learning process stopped after 37 epochs and that the decay of the MSE errors is quite fast, smooth and similar for both losses. 
The MSE errors go from $8.36 \times 10^{-2}$ to $2.23 \times 10^{-6}$ in 35 epochs for the training loss and from $1.39 \times 10^{-2}$ to $2.14 \times 10^{-6}$ for the validation loss. 
Moreover, the MSE errors do not move away from each other during training which indicates that the model was not overfitted. 

\subsubsection{Prediction of the ROM MLP}
\begin{figure}[hbtp!]
	\begin{subfigure}{0.5\textwidth}
		\centering
		\includegraphics[width=\textwidth]{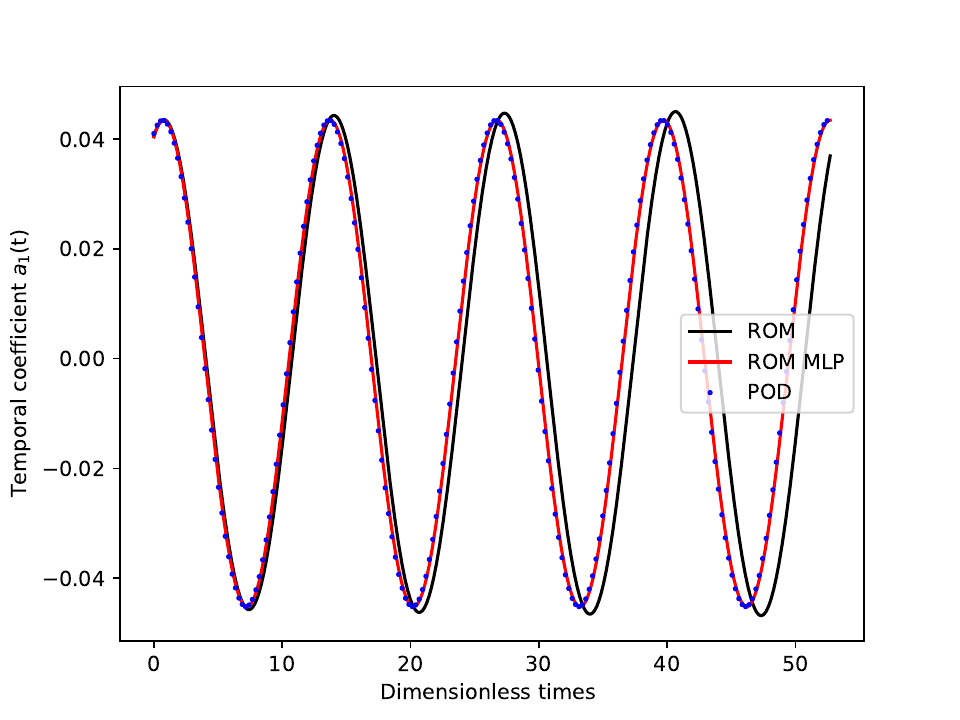}
		\caption{Mode 1}
	\end{subfigure}
	\hspace*{0.2cm}
	\begin{subfigure}{0.5\textwidth}
		\centering
		\includegraphics[width=\textwidth]{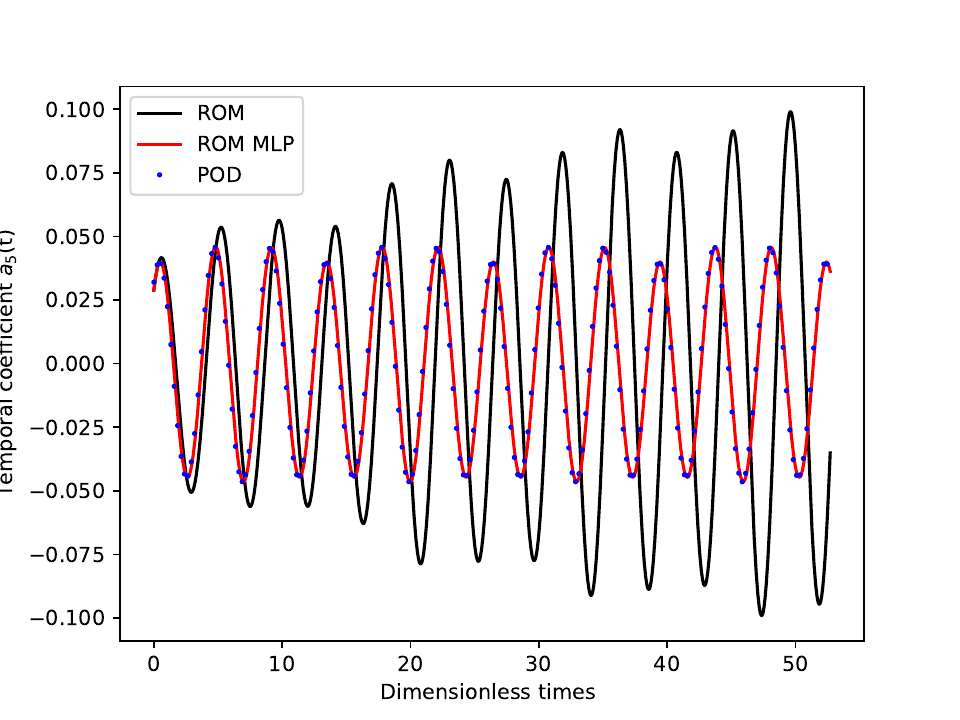}
		\caption{Mode 5}
	\end{subfigure}
	\begin{subfigure}{0.5\textwidth}
		\centering
		\includegraphics[width=\textwidth]{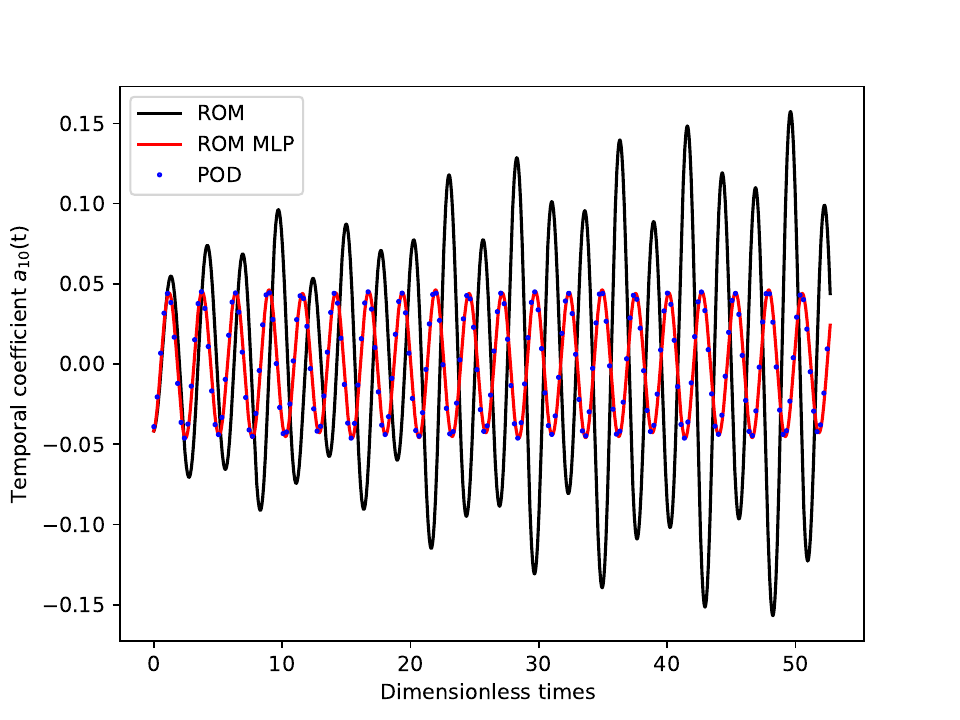}
		\caption{Mode 10}
	\end{subfigure}
	\hspace*{0.2cm}
	\begin{subfigure}{0.5\textwidth}
		\centering
		\includegraphics[width=\textwidth]{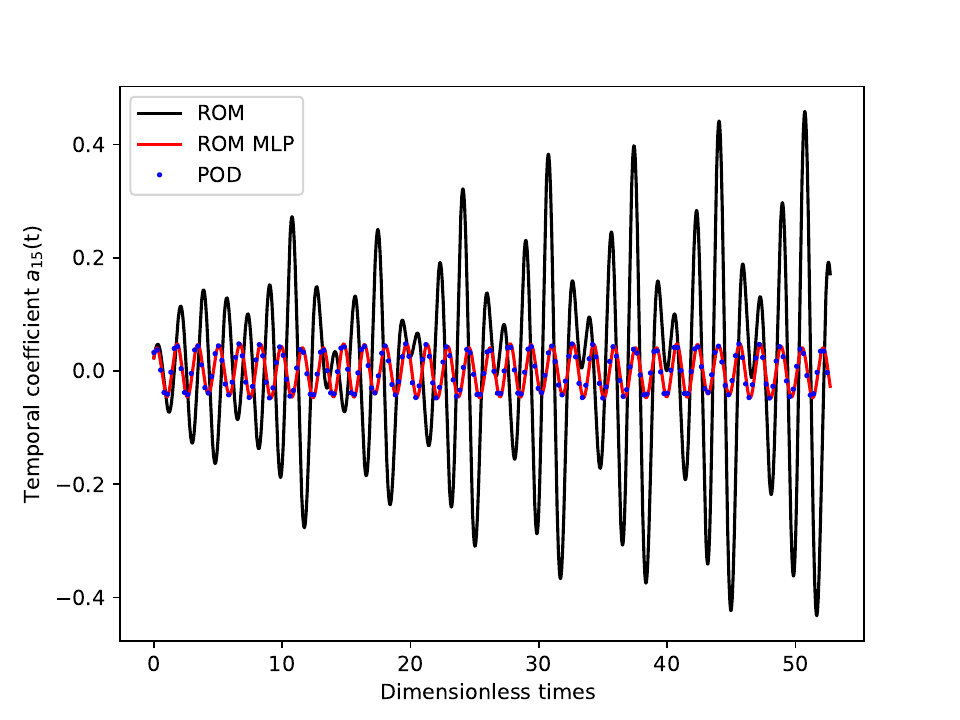}
		\caption{Mode 15}
	\end{subfigure}
	\caption{Temporal evolution of the coefficients $a_i(t)$ calculated during the sampling period for a Reynolds number $Re=110$. ROM corresponds to the reduced order model without residuals, ROM MLP corresponds to the reduced order model with the residuals calculated with MLP and POD is the reference. }
	\label{fig:unReyn_at}
\end{figure}

Next, calculations were performed with the reduced order model without residuals (or standard reduced order model) on the one hand, and with residuals learnt by the multilayer perceptron on the other hand (with the 1-layer model and 800 neurons, and 16 POD modes). In figure \ref{fig:unReyn_at} which represents the temporal evolution of the coefficients $a_i(t)$ during the sampling period, one can observe that the ROM MLP predicts the coefficients with a very good accuracy, and that the standard model does not yield satisfactory results. With the standard model, one can notice a small discrepancy for $a_1(t)$, and an increasing degradation for higher mode orders.\\

\begin{figure}[hbtp!]
	\begin{subfigure}{0.5\textwidth}
		\centering
		\includegraphics[width=\textwidth]{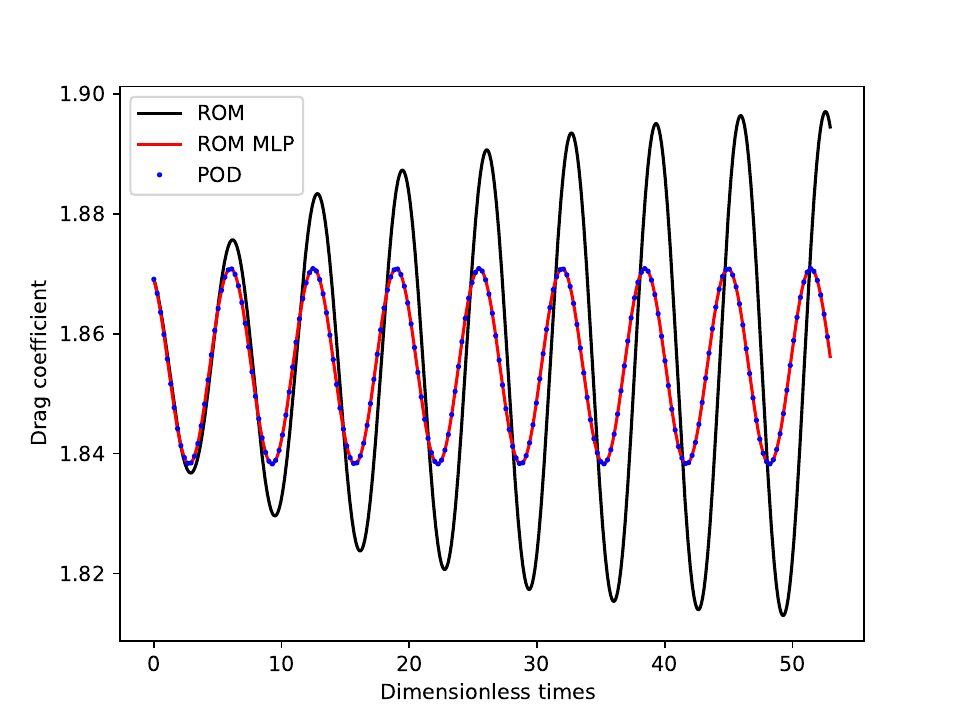}
		\caption{$C_d$}
	\end{subfigure}
	\hspace*{0.2cm}
	\begin{subfigure}{0.5\textwidth}
		\centering
		\includegraphics[width=\textwidth]{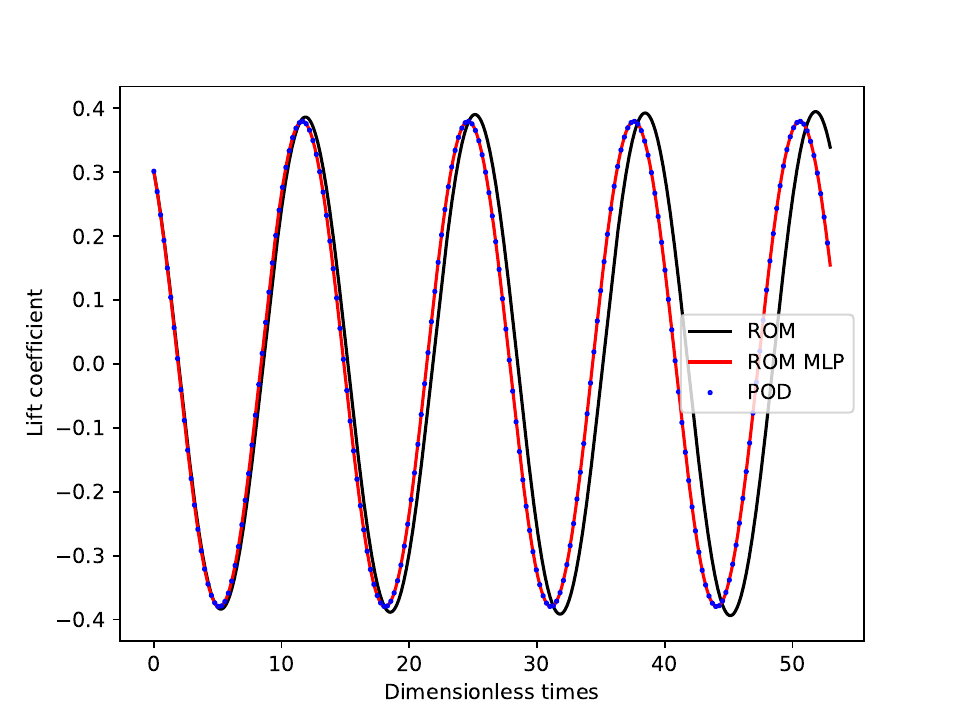}
		\caption{ $C_L$}
	\end{subfigure}
	\caption{Temporal evolution of the drag and lift coefficients calculated during the sampling period for a Reynolds number $Re=110$. ROM corresponds to the reduced order model without residuals, ROM MLP corresponds to the reduced order model with the residuals calculated with MLP and POD is the reference.}
	\label{fig:unReyn_cdcl_tempscourt}
\end{figure}

\begin{figure}[hbtp!]
	\begin{subfigure}{0.5\textwidth}
		\centering
		\includegraphics[width=\textwidth]{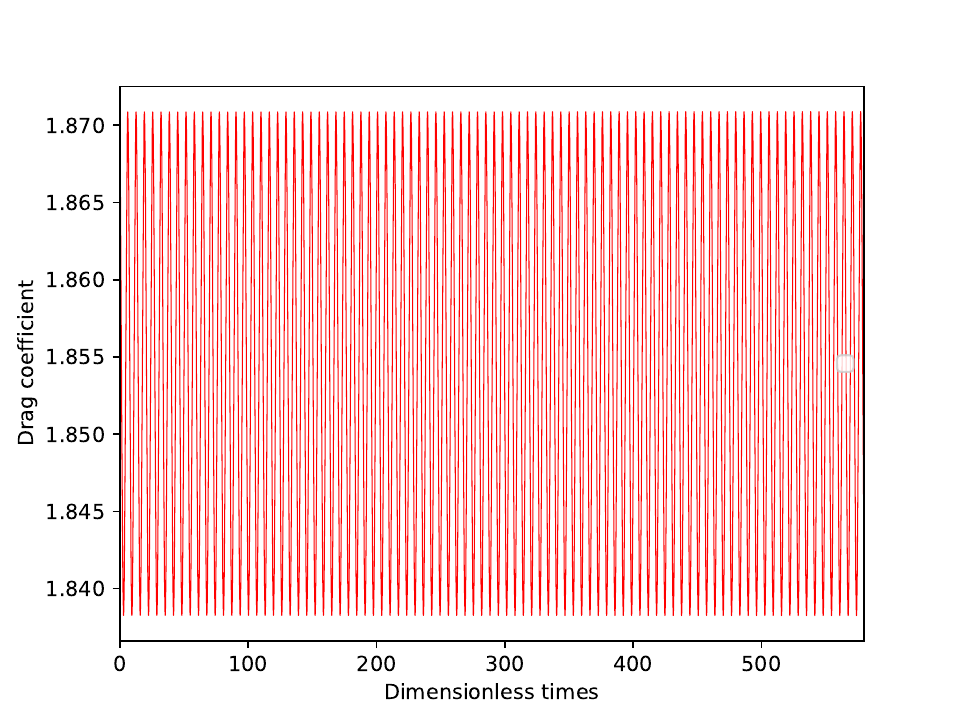}
		\caption{$C_d$}
	\end{subfigure}
	\hspace*{0.2cm}
	\begin{subfigure}{0.5\textwidth}
		\centering
		\includegraphics[width=\textwidth]{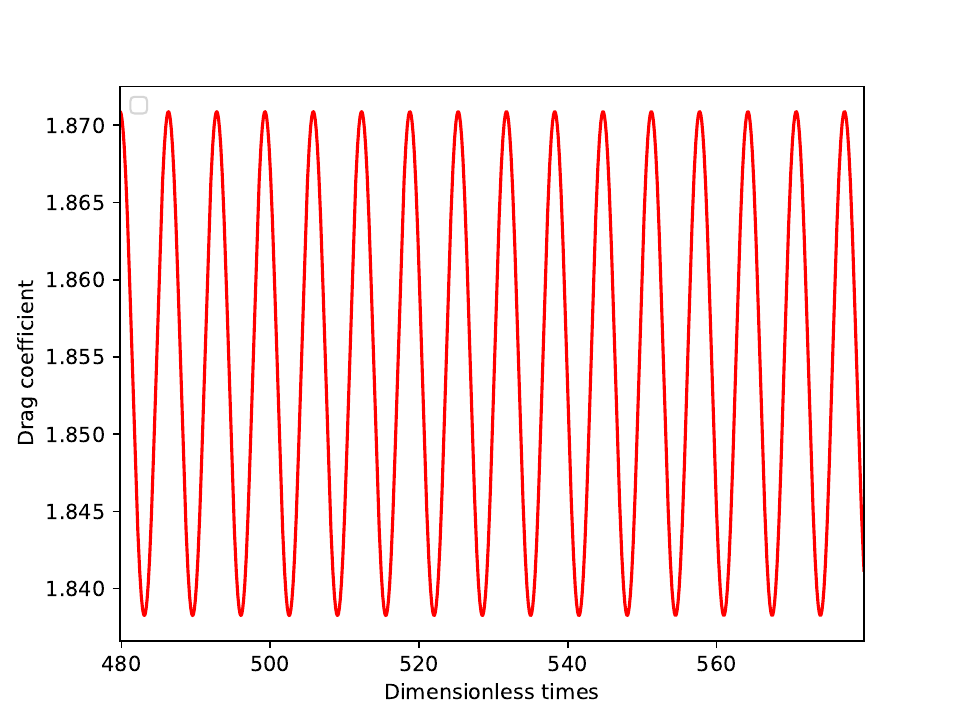}
		\caption{ $C_d$ (zoom on the last times)}
	\end{subfigure}
	\begin{subfigure}{0.5\textwidth}
		\centering
		\includegraphics[width=\textwidth]{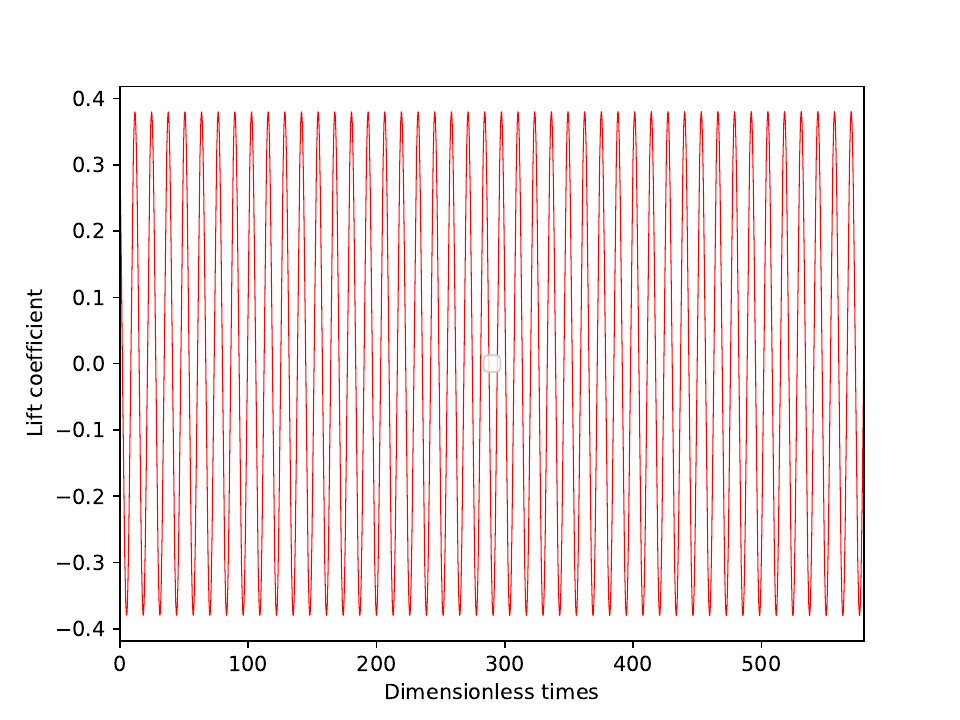}
		\caption{$C_L$}
	\end{subfigure}
	\hspace*{0.2cm}
	\begin{subfigure}{0.5\textwidth}
		\centering
		\includegraphics[width=\textwidth]{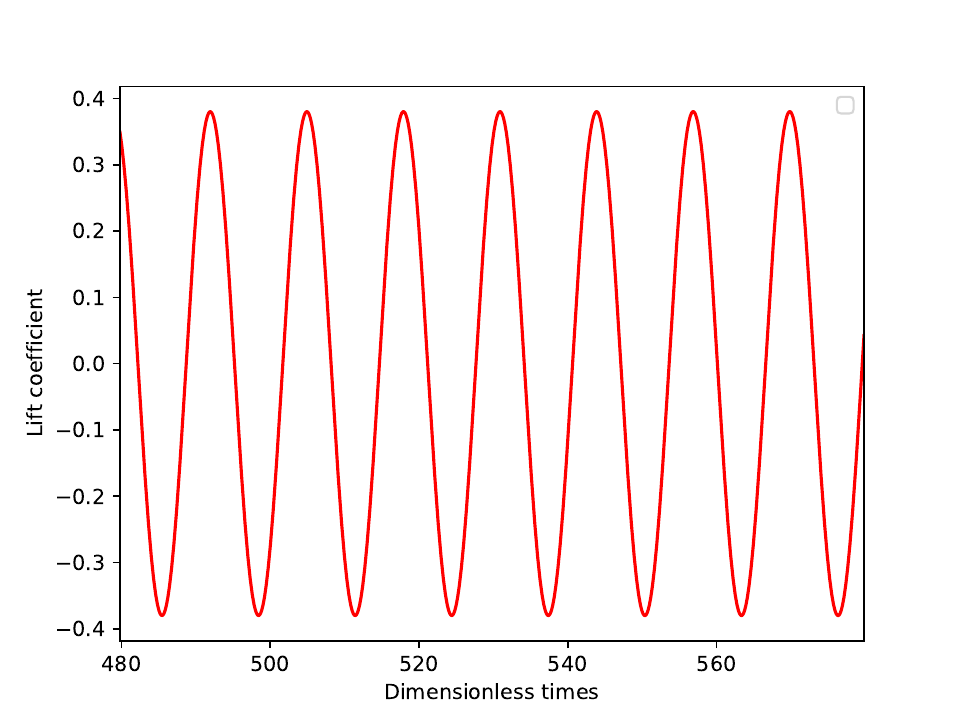}
		\caption{ $C_L$ (zoom on the last times)}
	\end{subfigure}
	\caption{Drag and lift coefficients obtained with the ROM MLP model for durations larger than the sampling period (Re=110).}
	\label{fig:unReyn_cdcl_tempslong}
\end{figure}

The temporal evolutions of the drag and lift coefficients during the sampling period are plotted in figure \ref{fig:unReyn_cdcl_tempscourt}. While the reduced order model without residuals is not able to reproduce the drag and lift coefficients during the sampling period, the reduced order model which calculates the residuals learnt by the MLP succeeds in predicting the drag and lift coefficients during the sampling period with an excellent accuracy. 
In addition, the behavior of the reduced order model with MLP for durations larger than the sampling period is satisfactory: a sinusoidal evolution of the coefficients according to time can be noted (see figure \ref{fig:unReyn_cdcl_tempslong}). The ROM MLP is very efficient and stable.

\subsection{Results obtained with the low order reduced models with and without residuals for three Reynolds numbers}
We now focus on including the Reynolds number in the features of the MLP model. 
The adaptation of the algorithm \ref{subsec:algorithm} to this case is quite obvious: 
the Reynolds number is simply added to the last column of feature matrix ${\bf A} \in \mathbb{R}^{N_s \times (N_u + 1)}$, where $N_s$ is the number of snapshots and $N_u$ is the number of selected modes.\\
\subsubsection{POD global basis}
\begin{figure}[hbtp!]
	\centerline{\includegraphics[width=0.75\textwidth]{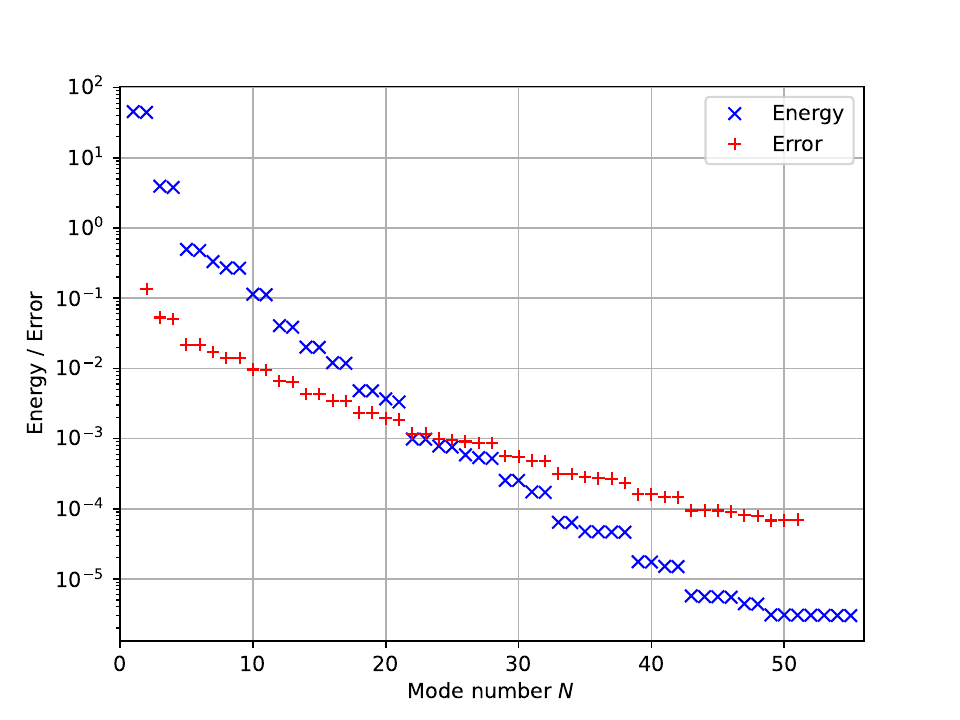}}
	\caption{Amount of fluctuating energy of each POD mode (blue curve) and maximum relative error between the solution obtained by the Navier-Stokes equations and the solution built by POD computed by keeping $N$ POD modes (red curve).}
	\label{fig:Multi_energy}
\end{figure}

\begin{figure}[hbtp!]
	\begin{subfigure}{0.5\textwidth}
		\centering
		\includegraphics[width=\textwidth]{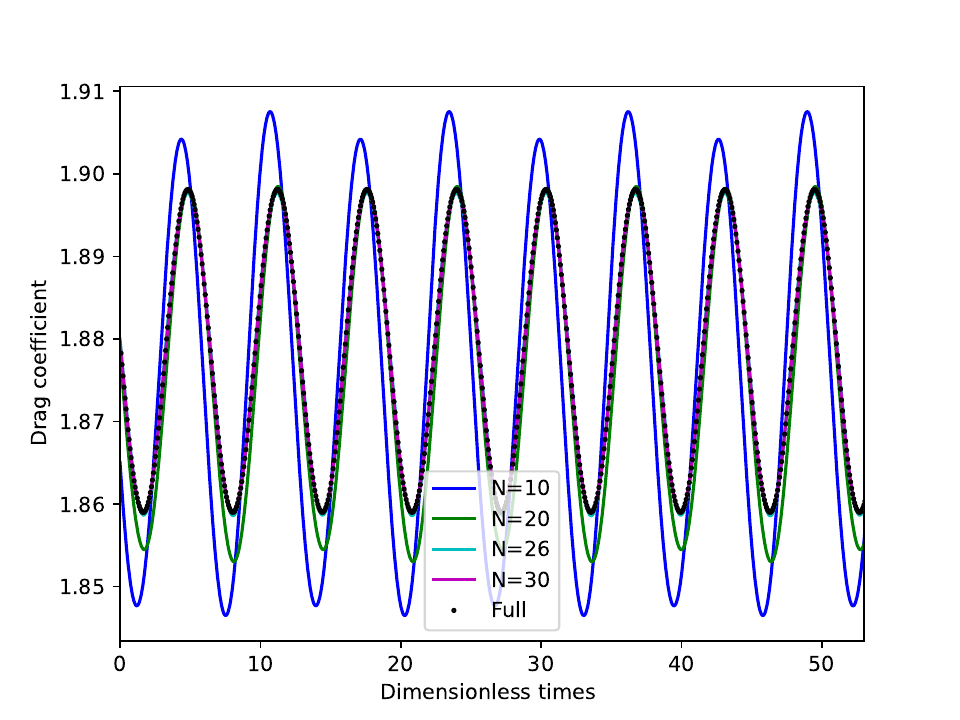}
		\caption{Re=120}
	\end{subfigure}
	\hspace*{0.2cm}
	\begin{subfigure}{0.5\textwidth}
		\centering
		\includegraphics[width=\textwidth]{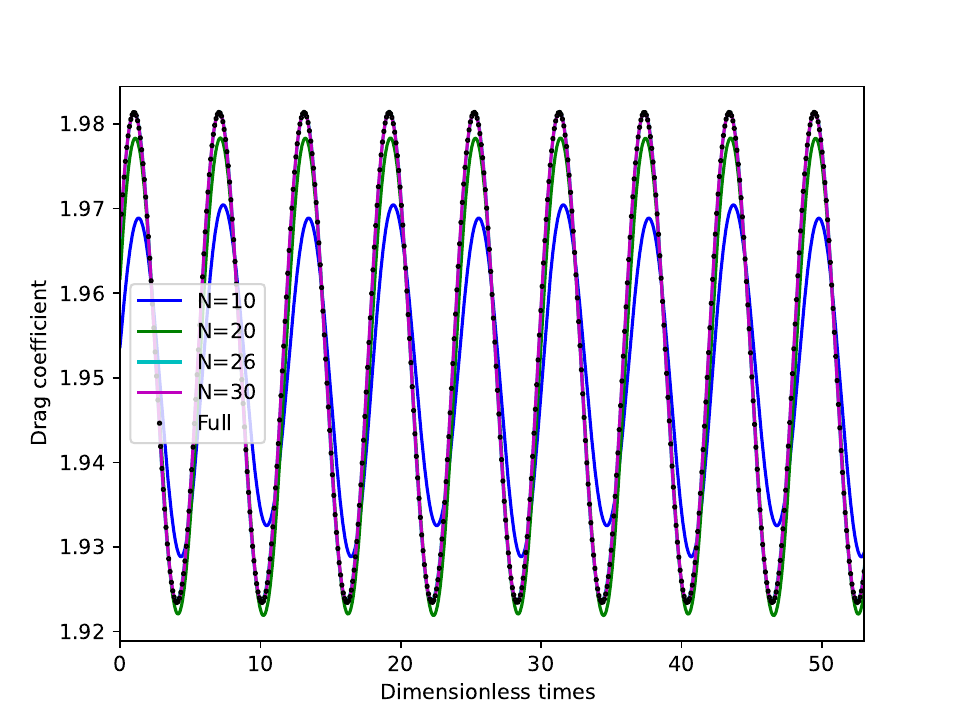}
		\caption{Re=160}
	\end{subfigure}
	\caption{Temporal evolutions of the drag coefficient for different numbers of the retained POD modes, calculated with the reference POD coefficients, for Re=120 and Re=160.}
	\label{fig:Multi_cd_modesPOD}
\end{figure}

\begin{figure}[hbtp!]
	\begin{subfigure}{0.5\textwidth}
		\centering
		\includegraphics[width=\textwidth]{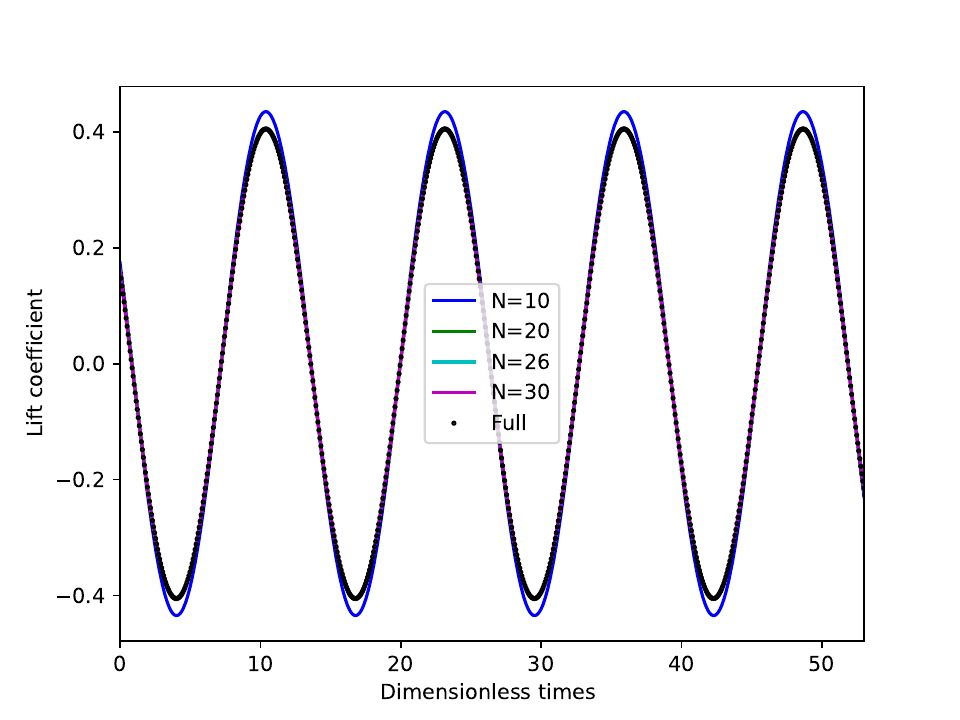}
		\caption{Re=120}
	\end{subfigure}
	\hspace*{0.2cm}
	\begin{subfigure}{0.5\textwidth}
		\centering
		\includegraphics[width=\textwidth]{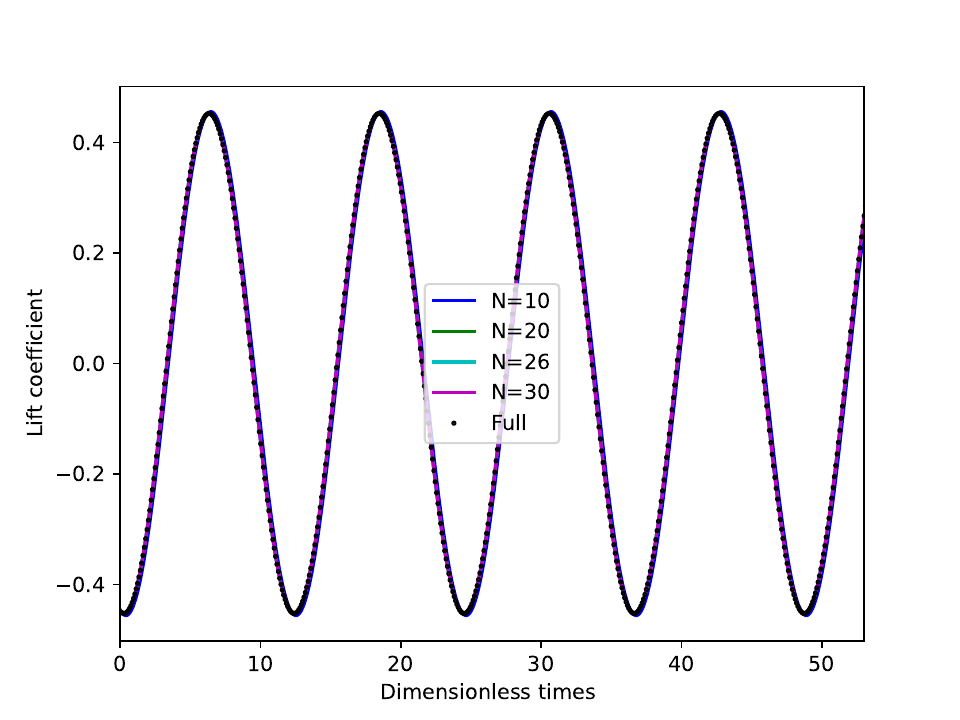}
		\caption{Re=160}
	\end{subfigure}
	\caption{Temporal evolutions of the lift coefficient for different numbers of the retained POD modes, calculated with the reference POD coefficients, for Re=120 and Re=160.}
	\label{fig:Multi_cl_modesPOD}
\end{figure}

In order to test the algorithm, a dataset of 3000 data (1000 data for each Reynolds number $\text{Re} \in \left\{ 120, 140, 160 \right\}$) was constituted and the corresponding eigenvalues and modes were calculated. To this end, a global correlation matrix was built. The eigenvalue problem (\ref{eigenvalue}) was solved, which led to the fluctuating energy presented in figure (\ref{fig:Multi_energy}). 
This figure shows that with 26 modes, the fluctuating energy was less than $10^{-3}$, and the maximum relative error between the solution of the Navier-Stokes equations and the velocity computed by POD was less than $10^{-3}$. In figures \ref{fig:Multi_cd_modesPOD} and \ref{fig:Multi_cl_modesPOD}, the temporal evolutions of the drag and lift coefficients calculated with the reference POD coefficients (see equation (\ref{aiPOD})) are presented according to the number of kept POD modes. It can be seen that the drag coefficient is very sensitive to the number of kept modes. 26 modes were then chosen to solve the low order dynamical system, with or without residuals calculated with MLP. \\

\subsubsection{Creation of the MLP model}
\begin{figure}[hbtp!]
	\centerline{\includegraphics[width=0.75\textwidth]{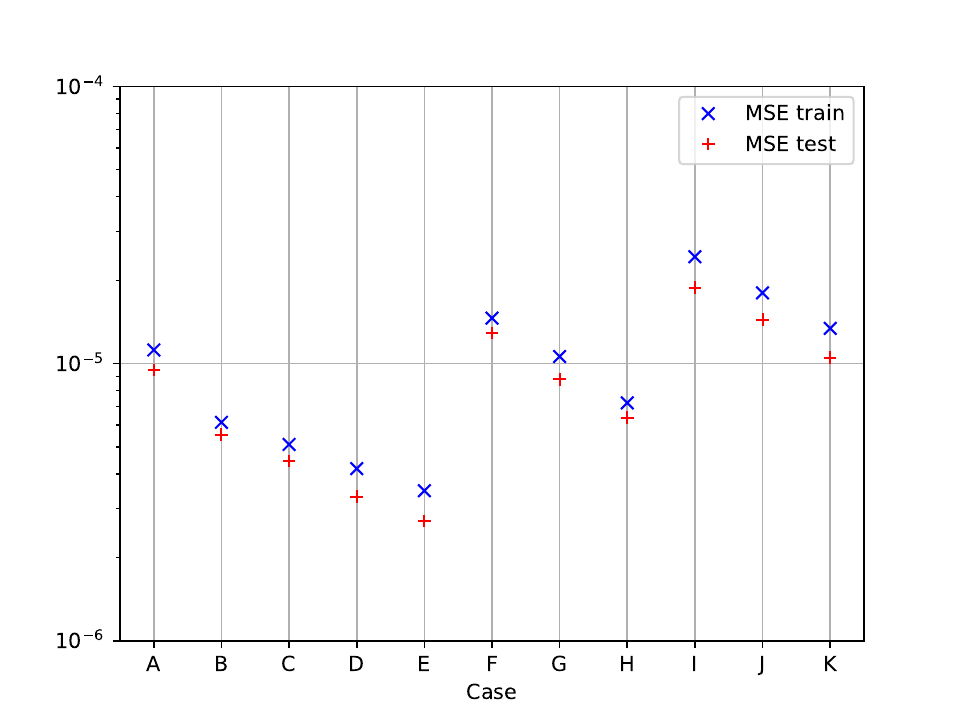}}
	\caption{Mean MSE errors over the validation sets for different values of the hyperparameters in the case of three Reynolds numbers and 16 modes.}
	\label{fig:Multi_msecase}
\end{figure}

For each Reynolds number $\text{Re} \in \left\{ 120, 140, 160 \right\}$, a dataset of 3000 data was built, which corresponded to the residuals ${\bf R}\in \mathbb{R}^{N_s\times N_u}$ associated with the temporal coefficients and each Reynolds number ${\bf A}\in \mathbb{R}^{N_s\times (N_u+1)}$. 
For each Reynolds number, the same amount of data was randomly split in the training, validation and test sets. For the 3000 data and the 26 modes, the residuals were learnt by the multilayer perceptron model. The same hyperparameter search as in the previous numerical study was carried out for $n_{\text{layers}} \in \left\{ 1, 2, 3 \right\}$ and a varying number of neurons (see table \ref{table:Multi_msecase} and figure \ref{fig:Multi_msecase} ). It is observed that the previous conclusions remain valid concerning the feature scaling method and the number of layers. Here again, the smallest errors are obtained for one layer. Moreover, the errors decrease for an increasing number of neurons. As a result, one layer and 2000 neurons were chosen for this study. In addition, the time needed to build the MLP model was 96s, and the computing time related to the ROM MLP was 161s, while the high fidelity computations required 2 hours and 40 minutes.\\

\begin{table}[hbtp!]
	\centering
	\csvreader[head to column names,tabular=|c|c|c|c|c|c|,
	table head=\hline \rowcolor{white!50} Case &  Nb layers&  Nb neurons &  Nb params &  MSE train &  MSE test ,
	late after head=\\\hline\rowcolor{darkgray!10},
	late after line=\csvifoddrow{\\\rowcolor{darkgray!10}}{\\\rowcolor{darkgray!25}}\hline]%
	{df_all_results_pour_latex2.csv}{Case=\Case,Nblayers=\Nblayers, Nbneurons=\Nbneurons,Nbparams=\Nbparams,MeanMSEtrain=\MeanMSEtrain,MeanMSEtest=\MeanMSEtest}%
	{\Case &\Nblayers & \Nbneurons  & \Nbparams  & \MeanMSEtrain  &  \MeanMSEtest  }
	\caption{Influence of the MLP model on the results obtained with the reduced order model.}
	\label{table:Multi_msecase}
\end{table}

Figure \ref{fig:unReyn_loss} shows the evolution of the training and validation losses with respect to the number of epochs. The model stagnates after 9 epochs. For this case, the learning process stopped after 9 epochs, and the MSE errors go from $1.82 \times 10^{-2}$ to $2.303 \times 10^{-6}$ in 9 epochs for the training loss and from $6.33 \times 10^{-4}$ to $4.42 \times 10^{-6}$ for the validation loss.\\

\begin{figure}[hbtp!]
	\centerline{\includegraphics[width=0.75\textwidth]{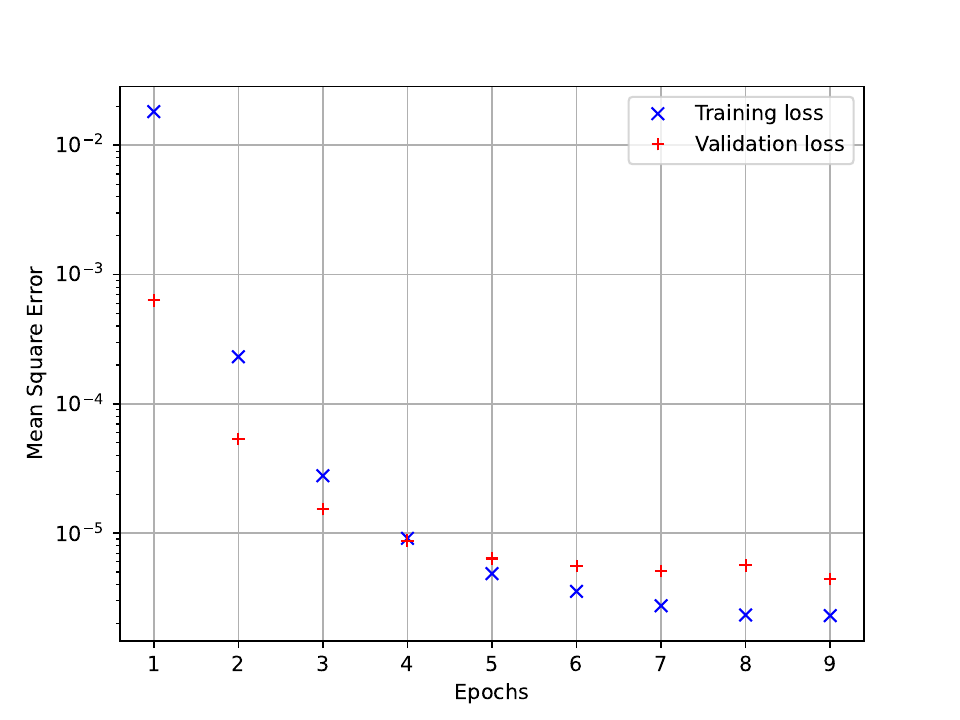}}
	\caption{Training and validation losses with respect to the number of epochs in the case of the global basis. The MSE is presented in logarithmic scales.
	}
	\label{fig:unReyn_loss}
\end{figure}

\subsubsection{Prediction of the ROM MLP}
\begin{figure}[hbtp!]
	\begin{subfigure}{0.5\textwidth}
		\centering
		\includegraphics[width=\textwidth]{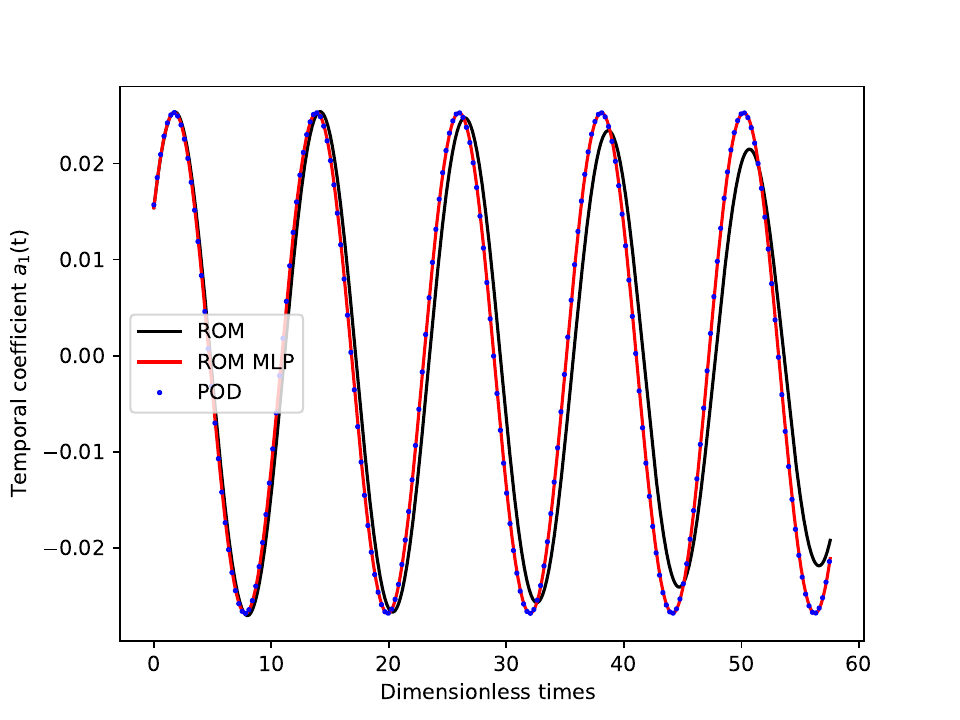}
		\caption{Mode 1}
	\end{subfigure}
	\hspace*{0.2cm}
	\begin{subfigure}{0.5\textwidth}
		\centering
		\includegraphics[width=\textwidth]{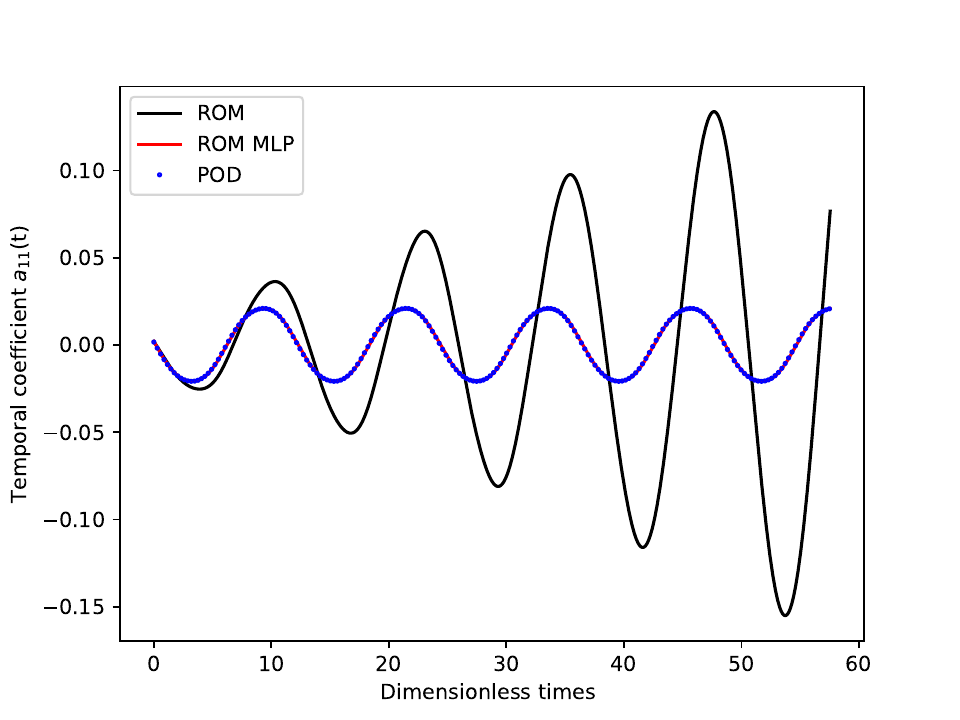}
		\caption{Mode 11}
	\end{subfigure}
	\begin{subfigure}{0.5\textwidth}
		\centering
		\includegraphics[width=\textwidth]{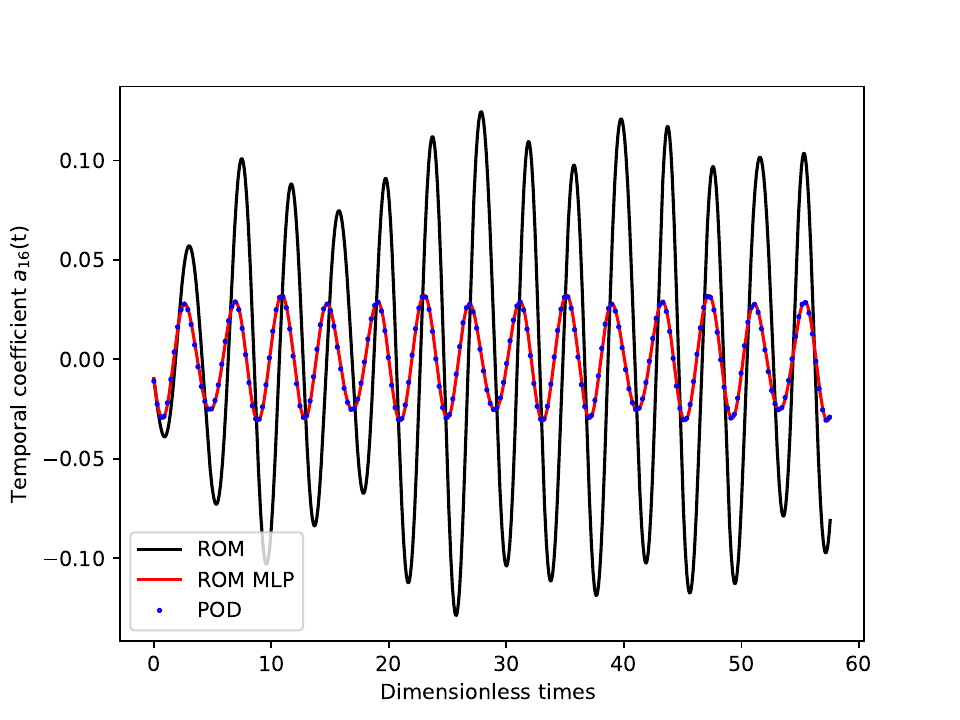}
		\caption{Mode 16}
	\end{subfigure}
	\hspace*{0.2cm}
	\begin{subfigure}{0.5\textwidth}
		\centering
		\includegraphics[width=\textwidth]{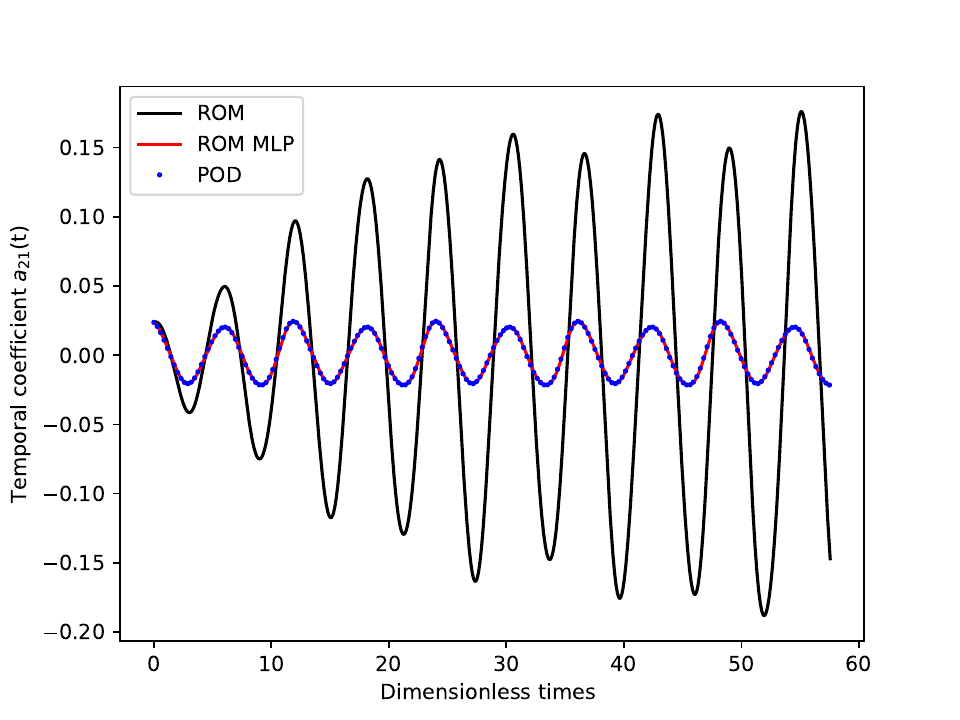}
		\caption{Mode 21}
	\end{subfigure}
	\caption{Temporal evolution of the coefficients $a_i(t)$ calculated during the sampling period for a Reynolds number $Re=160$. ROM corresponds to the reduced order model without residual, ROM MLP corresponds to the reduced order model with the residuals calculated with MLP and POD is the reference.}
	\label{fig:Multi_at_Re=160}
\end{figure}

\noindent Figure \ref{fig:Multi_at_Re=160} depicts the temporal evolution of the coefficients $a_i(t)$ for a Reynolds number $Re=160$. On the one hand, the reduced order model that learns the residuals with the multilayer perceptron predicts the  coefficients with a high accuracy for all modes, the coefficients predicted by the ROM MLP and the POD reference coefficients match each other. On the other hand, the reduced order model without residual fails to predict the coefficients, a small discrepancy between the coefficients calculated by the standard model and the POD reference coefficients can be noticed only for the first mode. This discrepancy increases for increasing POD modes. The results for Reynolds numbers equal to 120 and 140 are not shown but similar conclusions are effective.  

\begin{figure}[hbtp!]
	\begin{subfigure}{0.5\textwidth}
		\centering
		\includegraphics[width=\textwidth]{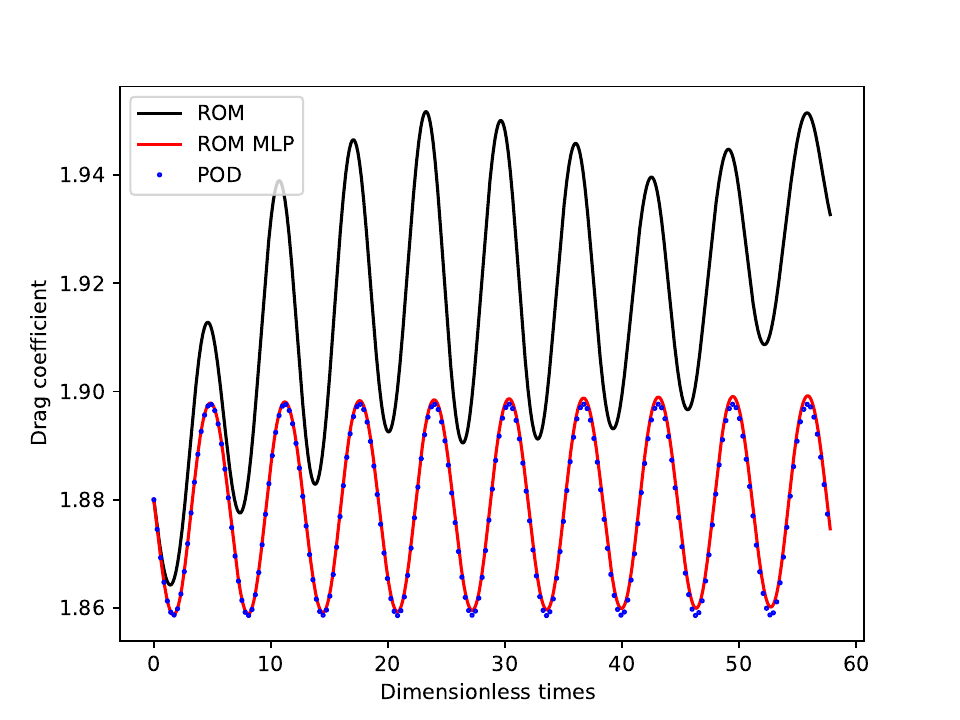}
		\caption{$C_d$ for Re=120}
	\end{subfigure}
	\hspace*{0.2cm}
	\begin{subfigure}{0.5\textwidth}
		\centering
		\includegraphics[width=\textwidth]{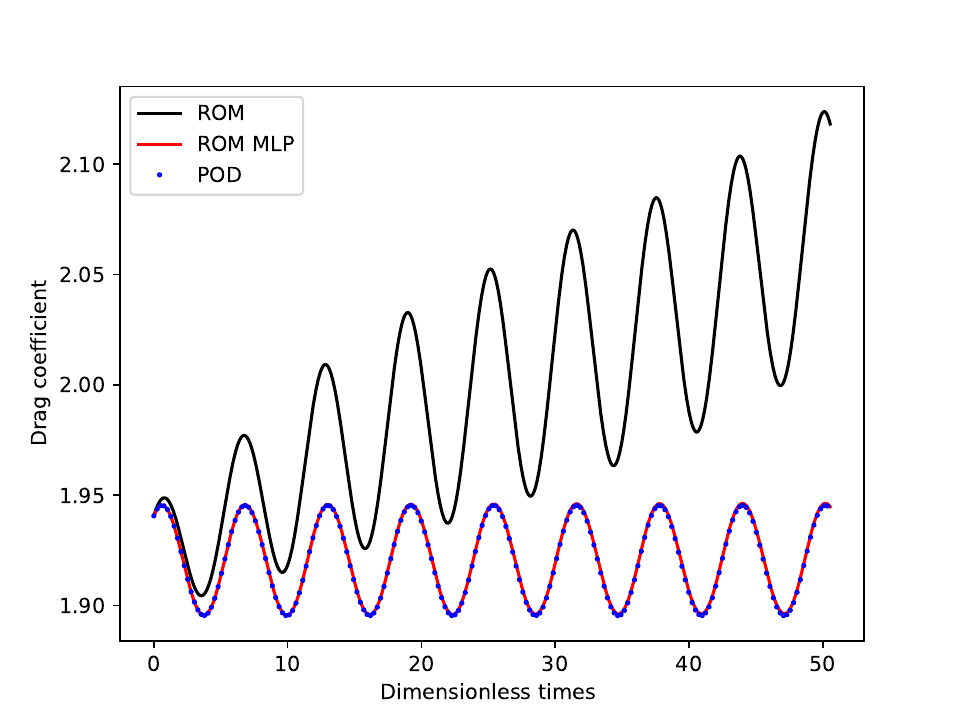}
		\caption{$C_d$ for Re=140}
	\end{subfigure}
	\begin{subfigure}{0.5\textwidth}
		\centering
		\includegraphics[width=\textwidth]{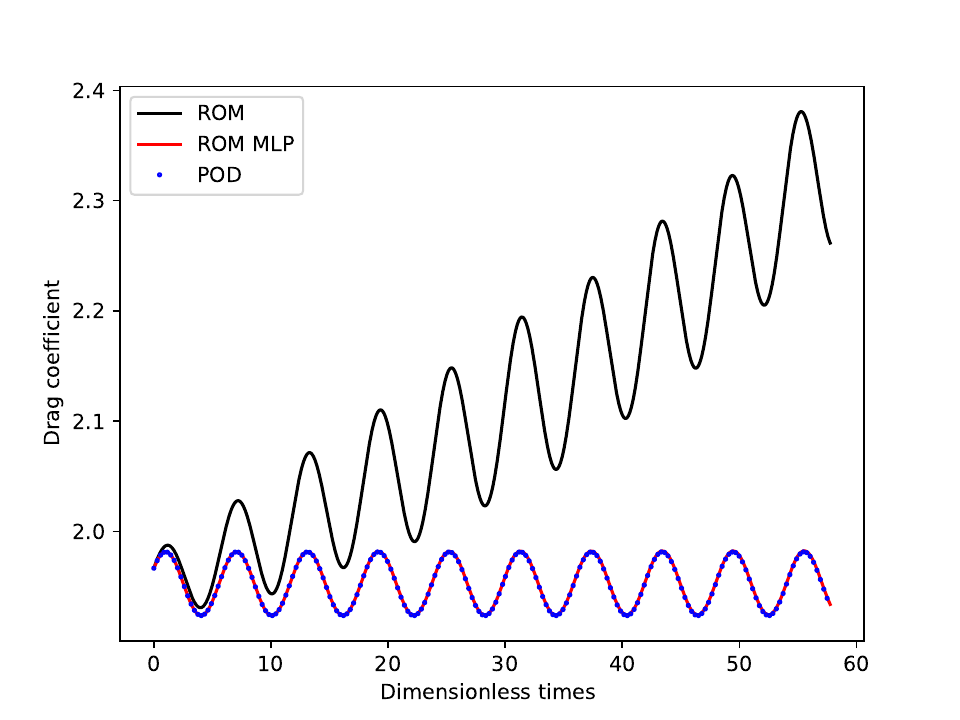}
		\caption{$C_d$ for Re=160}
	\end{subfigure}
	\hspace*{0.2cm}
	\begin{subfigure}{0.5\textwidth}
		\centering
		\includegraphics[width=\textwidth]{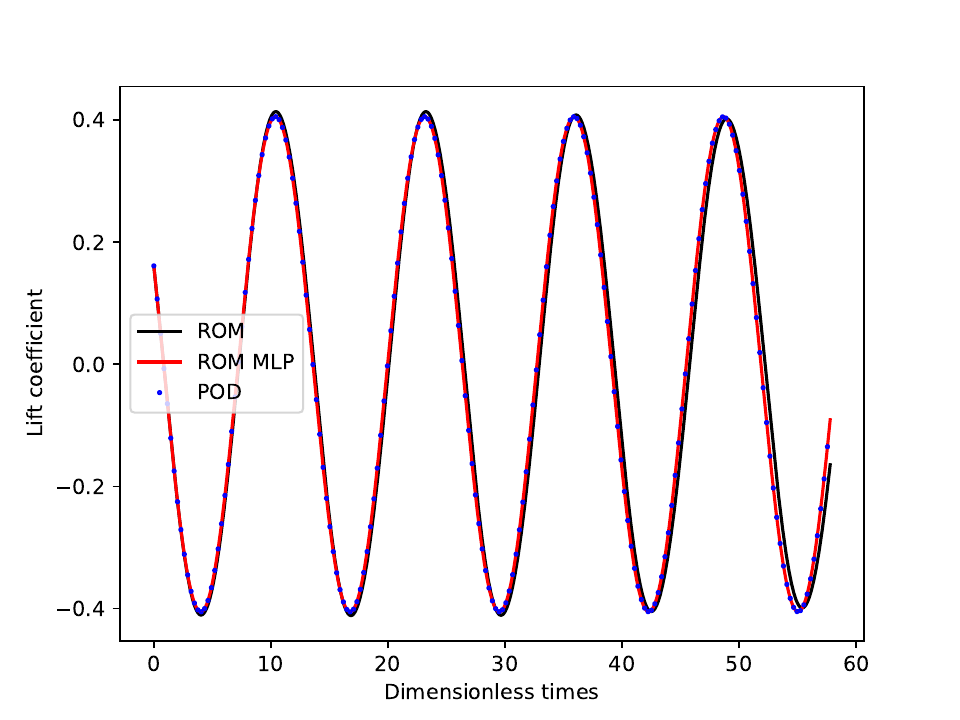}
		\caption{$C_L$ for Re=120}
	\end{subfigure}
	\begin{subfigure}{0.5\textwidth}
		\centering
		\includegraphics[width=\textwidth]{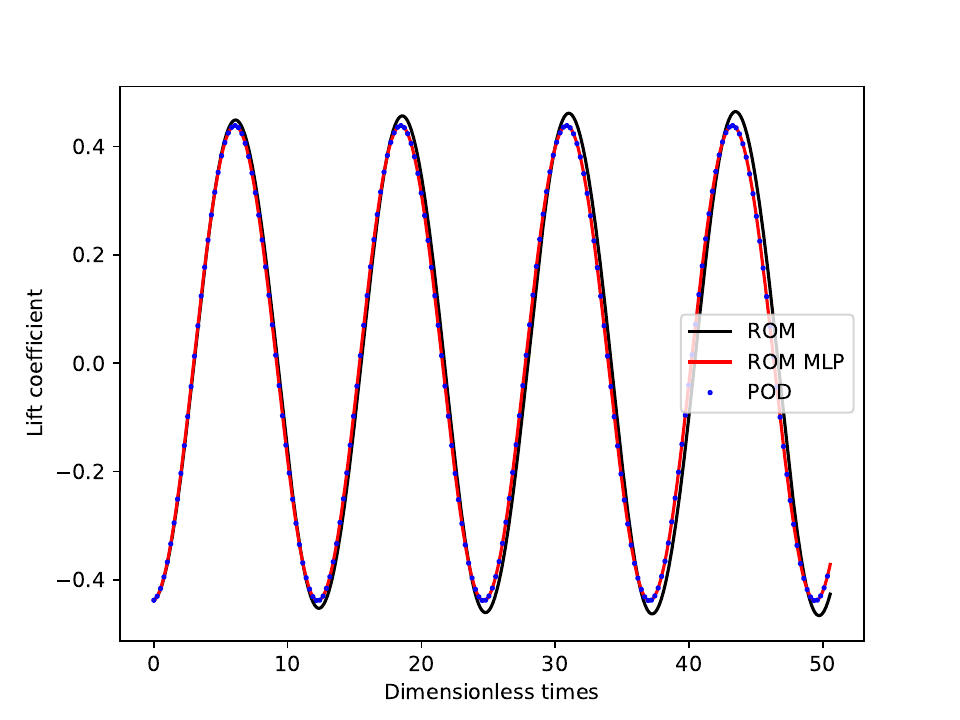}
		\caption{$C_L$ for Re=140}
	\end{subfigure}
	\hspace*{0.2cm}
	\begin{subfigure}{0.5\textwidth}
		\centering
		\includegraphics[width=\textwidth]{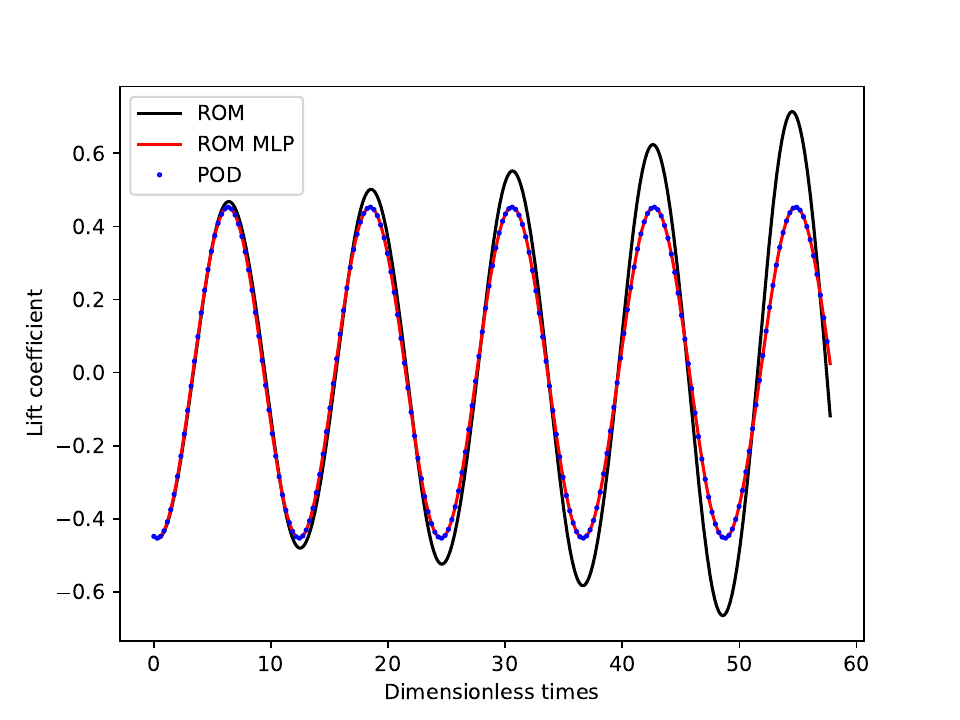}
		\caption{$C_L$ for Re=160}
	\end{subfigure}
	\caption{Temporal evolution of the drag and lift coefficients calculated during the sampling period for $\text{Re} \in \left\{ 120, 140, 160 \right\}$. ROM corresponds to the reduced order model without residual, ROM MLP corresponds to the reduced order model with the residuals calculated with MLP and POD is the reference. }
	\label{fig:Multi_cdcl_tempscourt}
\end{figure}

\begin{figure}[hbtp!]
	\begin{subfigure}{0.5\textwidth}
		\centering
		\includegraphics[width=\textwidth]{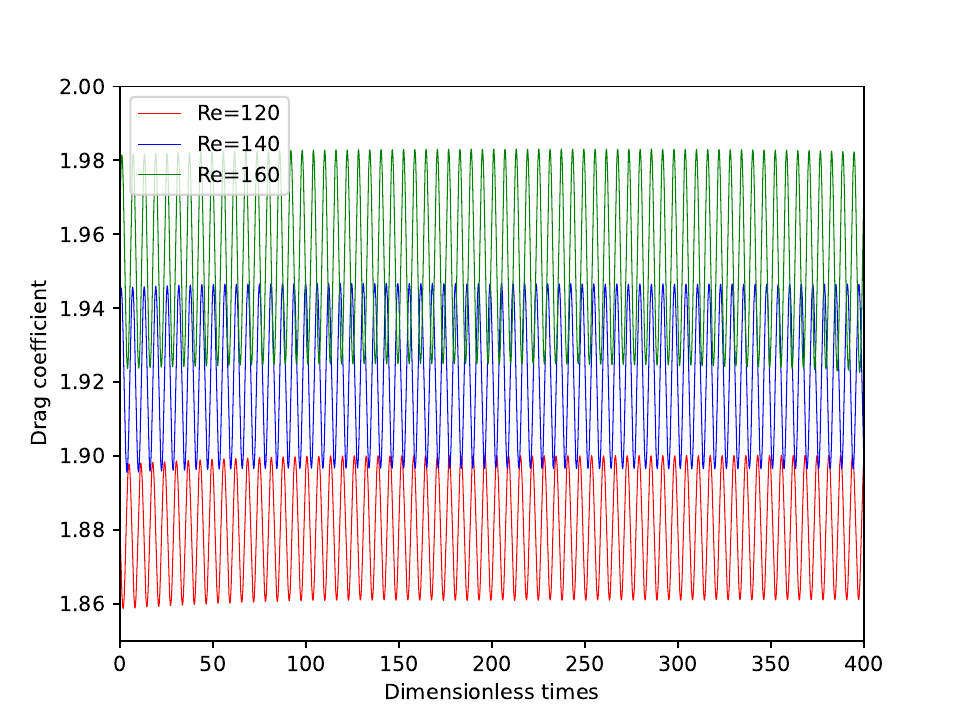}
		\caption{$C_d$}
	\end{subfigure}
	\hspace*{0.2cm}
	\begin{subfigure}{0.5\textwidth}
		\centering
		\includegraphics[width=\textwidth]{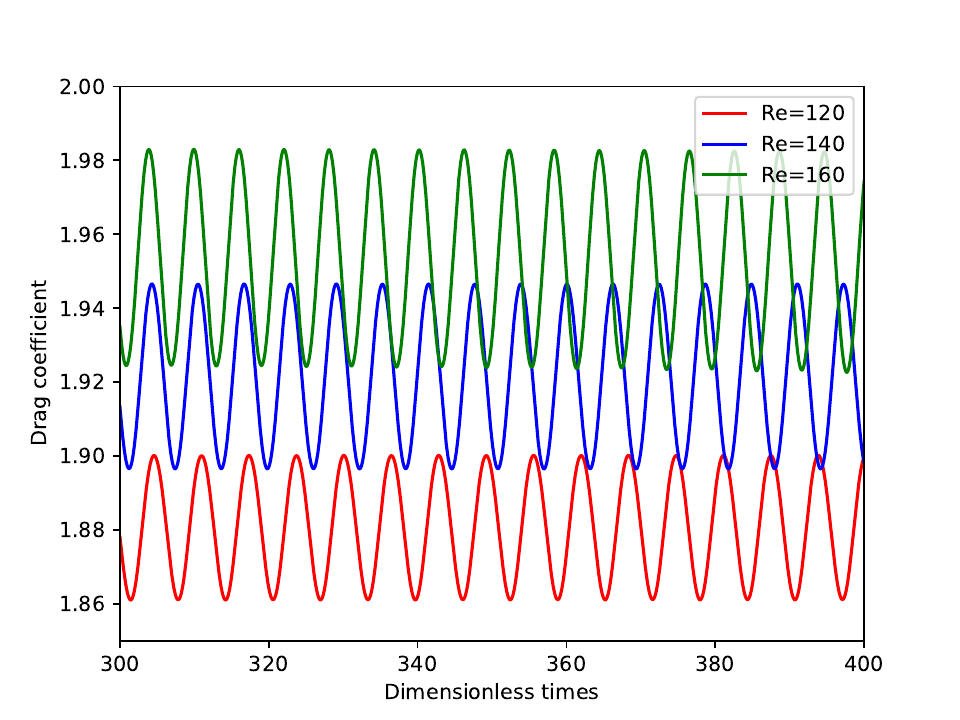}
		\caption{$C_d$ (zoom on the last times)}
	\end{subfigure}
	\begin{subfigure}{0.5\textwidth}
		\centering
		\includegraphics[width=\textwidth]{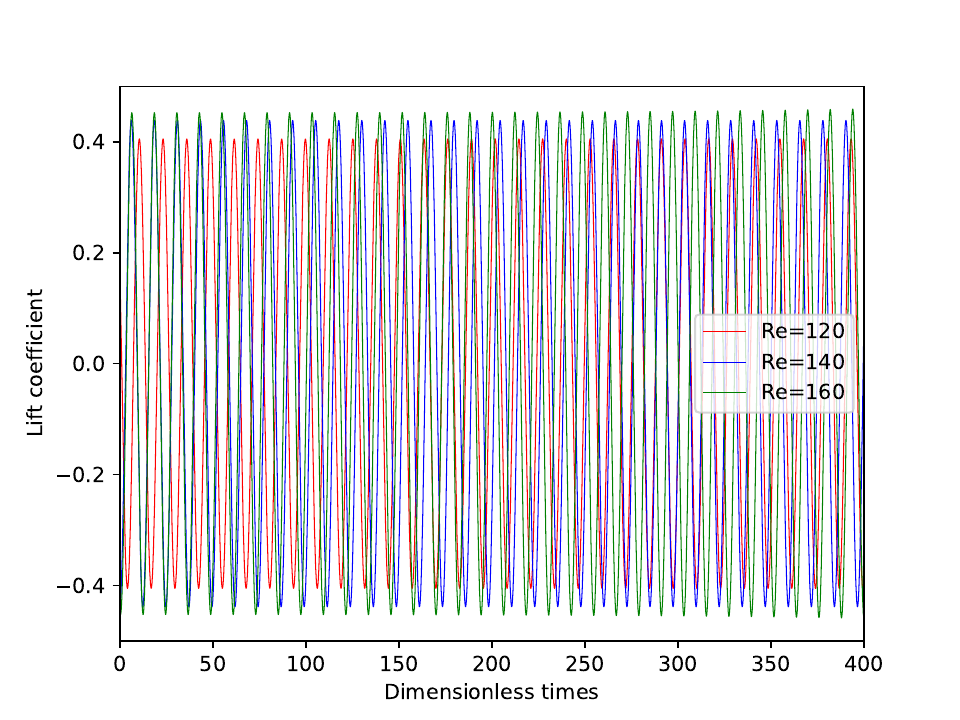}
		\caption{$C_L$}
	\end{subfigure}
	\hspace*{0.2cm}
	\begin{subfigure}{0.5\textwidth}
		\centering
		\includegraphics[width=\textwidth]{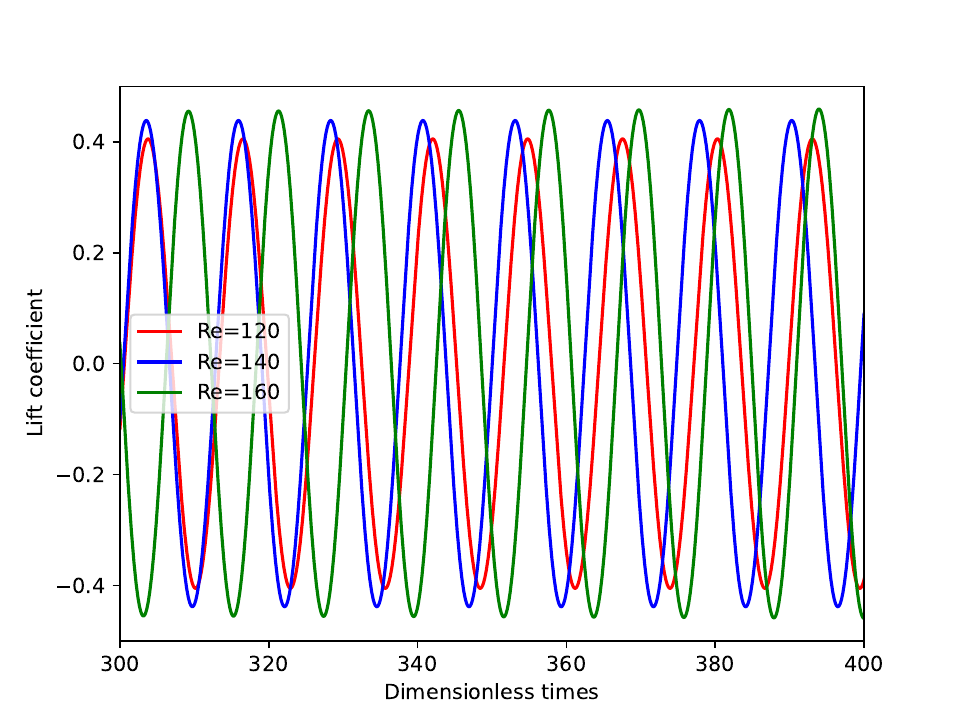}
		\caption{$C_L$ (zoom on the last times) }
	\end{subfigure}
	\caption{Temporal evolutions of the drag and lift coefficients predicted by the ROM MLP for durations larger than the sampling period, for $\text{Re} \in \left\{ 120, 140, 160 \right\}$.}
	\label{fig:Multi_cdcl_tempslong}
\end{figure}

In figure \ref{fig:Multi_cdcl_tempscourt}, the evolutions of the drag and lift coefficients during the sampling interval, calculated for $\text{Re} \in \left\{ 120, 140, 160 \right\}$ with the POD temporal coefficients (the reference values), the standard reduced order model (without residuals, ROM) and the reduced order model with the residuals calculated with MLP (ROM MLP), are presented. We can see that a very good agreement is obtained with the ROM MLP. The standard reduced order model yields wrong results, specially for the drag coefficient which increases with time. 
The evolutions of the drag and lift coefficients calculated with the ROM MLP for a duration much larger than the sampling time, are shown in figure \ref{fig:Multi_cdcl_tempslong}. This figure highlights that for times much larger than the sampling period, the temporal evolutions of the drag and lift coefficients predicted by the ROM MLP remain sinusoidal and stable.

\begin{figure}[hbtp!]
	\begin{subfigure}{0.5\textwidth}
		\centering
		\includegraphics[width=\textwidth]{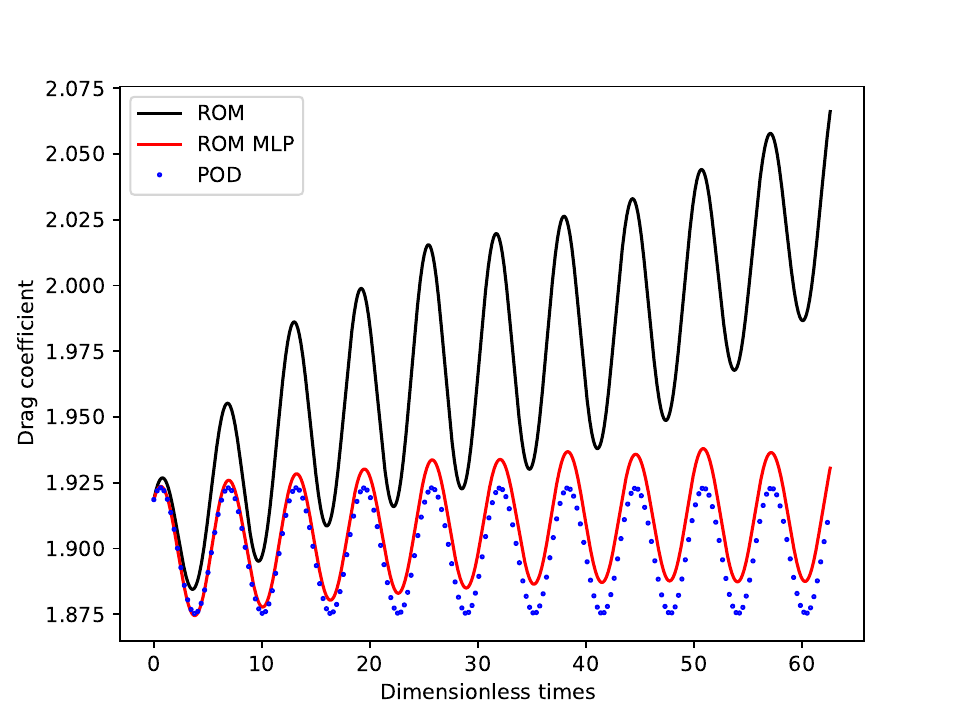}
		\caption{$C_d$ for Re=130}
	\end{subfigure}
	\hspace*{0.2cm}
	\begin{subfigure}{0.5\textwidth}
		\centering
		\includegraphics[width=\textwidth]{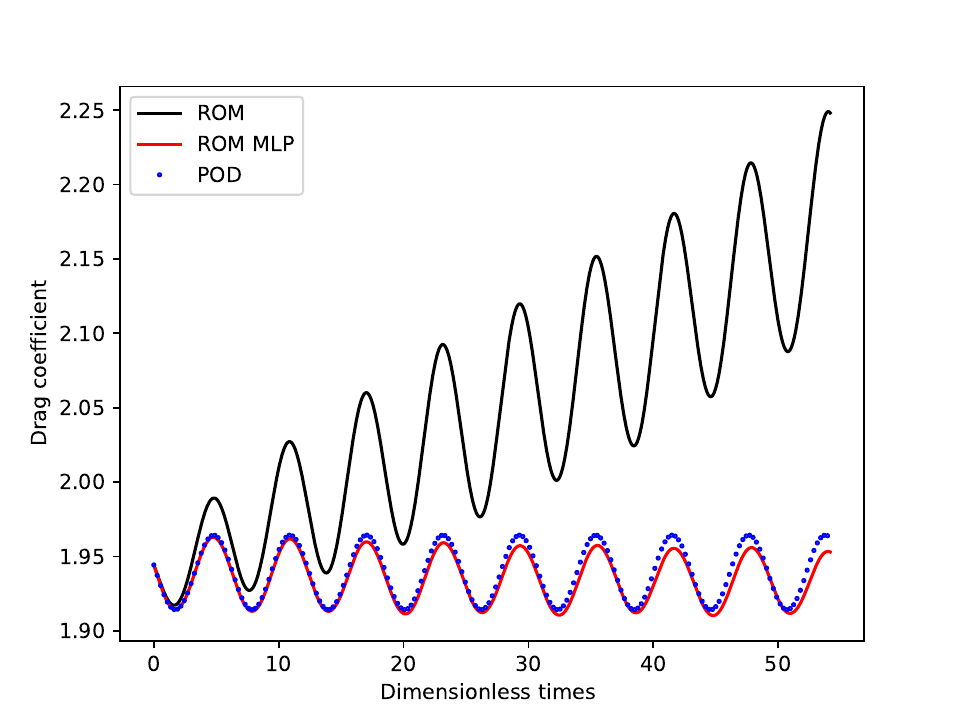}
		\caption{$C_d$ for Re=150}
	\end{subfigure}
	\begin{subfigure}{0.5\textwidth}
		\centering
		\includegraphics[width=\textwidth]{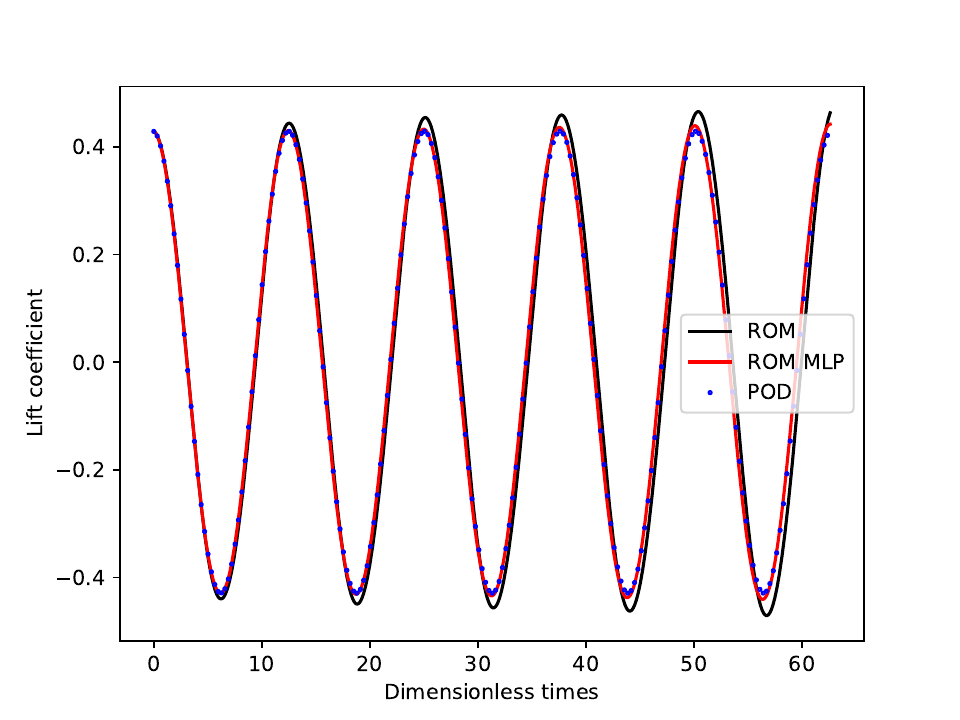}
		\caption{$C_L$ for Re=130}
	\end{subfigure}
	\hspace*{0.2cm}
	\begin{subfigure}{0.5\textwidth}
		\centering
		\includegraphics[width=\textwidth]{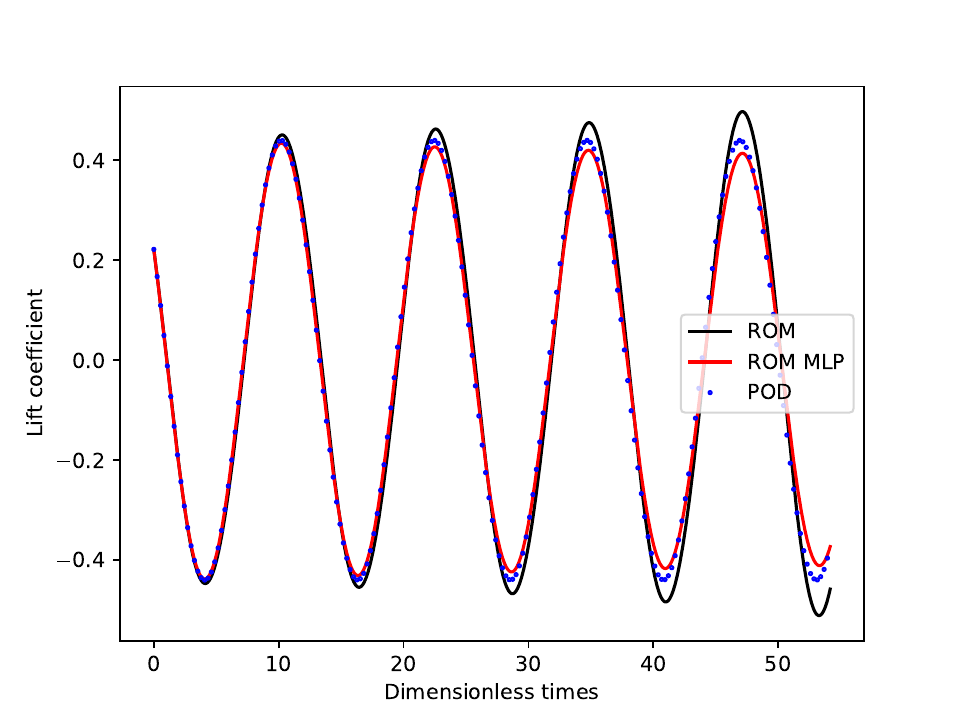}
		\caption{$C_L$ for Re=150}
	\end{subfigure}
	\caption{Temporal evolution of the drag and lift coefficients calculated during the sampling period for Reynolds numbers that were not used to build the POD basis and the MLP model. ROM corresponds to the reduced order model without residual, ROM MLP corresponds to the reduced order model with the residuals calculated with MLP and POD is the reference. }
	\label{fig:Multi_cdcl_tempscourt_prediction}
\end{figure}

Finally, the ROM MLP model was applied to $Re=130$ and $Re=150$. These Reynolds numbers were not used to build the POD basis and the MLP model. Figure \ref{fig:Multi_cdcl_tempscourt_prediction} shows the temporal evolution of the drag and lift coefficients during the sampling time. We can remark that the ROM MLP model yields results that are in good agreement with the reference values calculated with the POD coefficients, whereas a high discrepancy can be highlighted for the drag coefficient calculated with the reduced order model without residuals.

\section{Conclusions}
In this paper, a Galerkin POD reduced order model was proposed to calculate laminar flow around and through a porous obstacle in an unconfined channel. The non linear Forchheimer term, the influence of unresolved modes and pressure variations were taken into account by modeling the residuals of reduced order model with the multilayer perceptron method. This approach was validated by focusing on two cases : one case where the ROM  associated to multilayer perceptron method (ROM MLP) was applied to a single parameter (one Reynolds number) and one case where the ROM MLP was applied to a global POD basis built for three Reynolds numbers. For all the cases studied, the ROM MLP yielded satisfactory results while the reduced order model without residuals yielded wrong results. Moreover, the ROM
MLP method improves the prediction of flow for Reynols number that are not included
in the sampling and for times longer than sampling times. In addition, a substantial decrease of computing time in comparison with the high fidelity computations was highlighted.

\newpage

\bibliographystyle{plain}
\bibliography{bibliographie}

\end{document}